\documentclass[12pt]{JHEP3}
\usepackage[dvips]{epsfig}
\usepackage{cite}
\usepackage{verbatim}

\newlength{\wth}
\newcommand{\onegraph}[1]
{
\unitlength=1.1in
\begin{picture}(2.25,2.25)
 \put(-0.5,2.25){\epsfig{file=#1,width=0.65\wth,angle=270}}
\end{picture}
}

\newcommand{\onegraphwcntrs}[2]
{
\unitlength=1.1in
\begin{picture}(2.25,2.25)
 \put(-0.5,2.25){\epsfig{file=#1,width=0.65\wth,angle=270}}
 \put(-0.425,0.0){\epsfig{file=#2, width=0.915\wth}}
\end{picture}
}

\newcommand{\twographs}[2]
{ 
\unitlength=1.1in
\begin{picture}(5.4,2.25) 
 \put(-0.5,2.25){\epsfig{file=#1, width=0.65 \wth, angle=270}}
\put(2.3,2.25){\epsfig{file=#2, width=0.65 \wth, angle=270}}
\put(0.0,2.0){(a)}
\put(2.8,2.0){(b)}
\end{picture}
}

\newcommand{\mysixgraphs}[6]{%
 \unitlength=1.1in
 \begin{picture}(5.4,4.25)
 \put(-0.5,4.25){\epsfig{file=#1, width=0.65\wth, angle=270}}
 \put(2.3,4.25){\epsfig{file=#2, width=0.65\wth,angle=270}}
 \put(0.0,4.2){(a)}
 \put(2.8,4.2){(b)}
 \put(-0.5,2.0){\epsfig{file=#3, width=0.65\wth,angle=270}}
 \put(2.2,2.0){\epsfig{file=#4, width=0.65\wth,angle=270}}
 \put(0.0,1.95){(c)}
 \put(2.7,1.95){(d)}
\put(-0.425,2.0){\epsfig{file=#5,width=0.915\wth}}
\put(2.375,2.0){\epsfig{file=#6,width=0.915\wth}}
\end{picture}  
}

\newcommand{\twographswcntrs}[4]{%
 \unitlength=1.1in
 \begin{picture}(5.4,2.25)
 \put(-0.5,2.25){\epsfig{file=#1, width=0.65\wth, angle=270}}
 \put(2.3,2.25){\epsfig{file=#2, width=0.65\wth,angle=270}}
 \put(0.0,2.0){(a)}
 \put(2.8,2.0){(b)}
 \put(-0.425,0.0){\epsfig{file=#3, width=0.915\wth}}
 \put(2.375,0.0){\epsfig{file=#4, width=0.915\wth}}
\end{picture}  
}

\newcommand{\sevengraphs}[7]{%
\unitlength=1.1in
\begin{picture}(5.4,5.0)
\put(0.55,5.0){\epsfig{file=#1, width=0.45\wth, angle=270}}
\put(2.55,5.0){\epsfig{file=#2, width=0.45\wth,angle=270}}
\put(1.05,5.0){(a)}
\put(3.05,5.0){(b)}
\put(0.55,3.4){\epsfig{file=#3, width=0.45\wth,angle=270}}
\put(2.55,3.4){\epsfig{file=#4, width=0.465\wth,angle=270}}
\put(1.05,3.4){(c)}
\put(3.05,3.4){(d)}
\put(-0.5,1.5){\epsfig{file=#5, width=0.465\wth,angle=270}}
\put(1.4,1.5){\epsfig{file=#6, width=0.465\wth,angle=270}}
\put(3.3,1.5){\epsfig{file=#7, width=0.465\wth,angle=270}}
\put(0,1.5){(e)}
\put(1.9,1.5){(f)}
\put(3.8,1.5){(g)}
\end{picture}
}

\title{Global Fits of the Large Volume String Scenario to WMAP5 and Other Indirect Constraints Using Markov Chain Monte Carlo}

\author{Benjamin C Allanach$^{1}$, Matthew J Dolan$^{1}$ and Arne M
  Weber$^{2}$ \\ 
$^{1}$ DAMTP, CMS, Wilberforce Road, Cambridge CB3 0WA, UK\\
$^2$ Max Planck Inst.\ f\"{u}r Phys., F\"{o}hringer Ring 6, D-80805 Munich,
  Germany\\ 
}
\keywords{Supersymmetry Effective Theories, String Phenomenology}

\abstract{We present global fits to the Large Volume Scenario (LVS) of string models
using
current indirect data.
We use WMAP5 constraints on dark matter relic density, $b$-physics and
electroweak observables as well as direct search constraints. Such data can
be
adequately fit by LVS, with the best-fit point for $\mu>0$ having
$\chi^2=13.6$ for 8 degrees of freedom.
The resulting constraints on parameter space are robust in that they do not
depend much upon the prior, or upon whether one uses Bayesian or frequentist
interpretations of the data. Sparticle masses are constrained to be well
below
the 1 TeV level, predicting early SUSY discovery at the LHC.
We devise a method of quantifying which are the
most important constraints. We find that the LEP2 Higgs mass constraint, the
relic density of dark matter and the anomalous magnetic moment of the muon
affect the fits to the strongest degree.
}

\preprint{DAMTP-2008-37}

\begin{document}
\tableofcontents
\section{Introduction}
From the host of possibilities of Beyond the Standard Model physics 
the most well-studied and well-motivated is string theory. Stringy
models naturally incorporate supersymmetry (SUSY), which can solve the problems of the instability of the Higgs mass (the
hierarchy problem) and the nature of dark matter. Unfortunately the
simplest extension of the Standard Model which includes SUSY, the
so-called Minimal Supersymmetric Standard Model (MSSM) contains around
120 free parameters making a predictive analysis extremely
difficult. While the effective dimensionality of the parameter space
is reduced by constraints from flavour changing neutral currents, it
is still large enough to tax even the most powerful CPUs and
techniques available today. It is to be hoped that some organising
principle will be found at the LHC or a future collider to enable us
to understand the relationships between these parameters, but in the
meantime one popular way of dealing with this problem is by unifying
the scalar mass terms to $m_0$, the trilinear terms to $A$ and the
gaugino masses to $M$ at some high energy scale, usually taken to be
the GUT scale $\approx 10^{16}$ GeV. With this pattern of soft SUSY breaking terms the MSSM is known as the Constrained MSSM (CMSSM) or minimal supergravity (mSUGRA). While universality is \textit{a priori} a very strong assumption, there are several string motivated models which predict such universality \cite{unistring,LVS} and, more pragmatically, it renders the problem of performing a phenomenological analysis of the MSSM practical in computational terms.

One such string theoretic model is the Large Volume Scenario (LVS)\cite{LVS}in the limit of dilute fluxes. Originally discovered in the context of type IIB flux compactifications, these models achieve 4D $\mathcal{N}=1$ broken supersymmetry with all moduli stabilised and exponentially large compactification volume. All of these features are phenomenologically desirable. Having stabilised moduli means massless scalar particles and non-realistic fifth forces are avoided. Exponentially large volume gives us confidence that the working in the supergravity limit is viable. Furthermore, the large volume $\mathcal{V}$ lowers the string scale and gravitino mass to 
\begin{equation}
\label{eq:ms}
 m_s \sim \frac{M_P}{\sqrt{\mathcal{V}}}, \qquad m_{3/2} \sim \frac{M_P}{\mathcal{V}}.
\end{equation}
From eq.~(\ref{eq:ms}) it is clear that a volume $\mathcal{V}\sim
10^{16}$ in string units will lead to TeV scale supersymmetry breaking
as parametrised by the gravitino mass and to an intermediate string
scale $m_s\sim 10^{11}$ GeV. The benefits of an intermediate string
scale have previously been discussed in ref.\cite{intermediate} and
include a natural solution to the strong CP problem, the correct scale
of suppression of neutrino masses and gauge coupling
unification at $m_s$, a possibility not usually considered in the
context of mSUGRA. One further notable feature of the large volume
models is that the flux superpotential $W_0$ does not need to be fine
tuned and is naturally of $\mathcal{O}(1)$, unlike the case of the
well-known KKLT vacua \cite{KKLT} which require $W_0 \sim
10^{-13}$. The Large Volume Scenario is therefore one of the most well
studied and robust models which string theory presents. In this paper we shall confront the LVS with current indirect data from cosmology and particle physics, providing for the first time a global fit to the model where it is possible to trade goodness of fit between parameters, so that a bad fit to one observable may be compensated for by good fit to another.

The connection between the high scale supergravity F-terms and the MSSM soft terms was established in \cite{LVSsoftterms} and demonstrated soft-term universality in the limit of dilute fluxes. We have no knowledge of what the exact values of the high scale boundary conditions are, and a systematic exploration of the parameter space has yet to be performed. Such an exploration would tell us what the viable regions of parameter space are and equally as important, what regions have already been ruled out. We also wish to take into account uncertainties and errors in our knowledge and predictions about standard model parameters. The usual procedure of fixed grid scans through parameter space is therefore not useful, as well as being computationally intensive since the number of points scanned and hence the time taken is proportional to $k^N$, where $N$ is the number of important free parameters the model has and $k$ is the desired number of points along each dimension. In standard mSUGRA this is taken to be 8 and in our model it is 6. A more efficient method that allows us to take errors and uncertainties into account is the use of Monte Carlo Markov chains (MCMC) in a Bayesian statistical formalism, as was first considered in ref.\cite{Allanach:2005kz} and further developed in  refs.\cite{Allanach:2005kz, naturalness,deAustri:2006pe,leszek}.

What we are interested in is the posterior probability distribution function (pdf) $p(m|\textup{data})$, the probability of a point $m$ in parameter space being ``correct'' given some Standard Model data such as masses and other observables. Unfortunately is difficult to calculate: given a value of some branching ratio it is arduous to invert this to find which model parameters are consistent with it. However, it is quite straightforward to obtain the likelihood distribution $p(\textup{data}|m)$ which is the probability of obtaining some particular SM observables given a point in parameter space. Once we have the spectrum of the model at that point it is trivial to calculate things like branching ratios. The connection between the likelihood distribution and the posterior pdf has been provided for us by the Reverend Thomas Bayes and his eponymous theorem \cite{bayes}
\begin{equation}  \frac{p(m_1|\mbox{data})}{p(m_2|\mbox{data})} = \frac{p(\mbox{data}|m_1)p(m_1)}{p(\mbox{data}|m_2)p(m_2)}  \end{equation}
where $p(m_i)$ is known as the prior (pdf) or simply prior. It encodes our previous beliefs or uncertainties about a particular point in parameter space. In the case of flat priors where $p(m_1)=p(m_2)$  the likelihood distribution is proportional to the posterior pdf. We wish then to numerically construct the likelihood distribution. To do this we use Monte Carlo Markov chains. A Markov chain is best described as a string of points which sample from some continuous distribution. These have the benefit that the run time depends only linearly on the number of dimensions in parameter space, in contrast with the power law behaviour for grid scans. For more details on the implementation of the MCMC method we use in this paper see ref.\cite{Allanach:2005kz}. To construct the likelihood we will use a variety of data from cosmology, electroweak precision observables, $b$-physics and current sparticle mass limits, including for the first time in a Bayesian context $BR(B\to \tau\nu)$, the meson mass splitting $\Delta M_{B_s}$ and the isospin asymmetry $\Delta_{0-}$. We also make use of the new WMAP5 dataset and the most recent measured value of the top quark mass $m_t$ from the Tevatron.

Previous phenomenological studies of the LVS are refs.\cite{LVSLHC,LVSsoftterms} where random sample spectra were generated which were then used to place bounds on the viable regions of parameter space and investigate LHC collider observables and signatures of the model. This paper does not address the issue of collider observables, but extends the above cited work to accurately fit and sample the parameter space of the model using currently available indirect constraints.

Recently ref.\cite{weiglein} have performed a $\chi^2$ analysis of the CMSSM and minimal GMSB and AMSB scenarios using a subset of electroweak and $b$-physics observables. Their fits consist of random scans of parameter space using $10^5$ points. Aside from the fact by incorporating variations in the SM parameters and, importantly, the dark matter relic density which is known to be the strongest constraint on the CMSSM parameter space, our MCMC approach also allows for better sampling and hence statistically more stable results. On the other hand, we do not do multi-model hypothesis testing, but instead perform a hypothesis test on the sign of $\mu$, the bi-linear parameter of the Higgs potential whose magnitude is accurately known by fitting to $M_Z$.

In the next section we discuss the origin of the soft SUSY breaking terms in the Large Volume Scenario and some caveats regarding some approximations we make. In Section 3 we present our suite of observables with which we will constrain the model and  describe how we construct the likelihood, the issue of priors and how we test for convergence of our Markov chains. Section 4 presents our fits to the likelihood distribution and posterior pdfs for some sparticles and the dark matter relic density. We also discuss channels of relic density annihilation, best-fit points for the model and present a new variable constructed to make quantitative statements about which observables are constraining the form of the likelihood distribution the most. Finally, we entertain the possibility that $\mu<0$. Section 5 describes a frequentist's approach using profile likelihoods which we extract from our Markov chains and identifies some interesting `volume effects'. We conclude by recapitulating our main points and presenting some possible directions for future research.

\section{Soft terms in the LVS}

The Large Volume Scenario\cite{LVS} is one of the most successful
paradigms in string theory which achieves realistic low energy physics
with stabilisation of all moduli fields. From within this scenario one
can obtain TeV scale supersymmetry (SUSY) breaking,
inflation\cite{inflation}, QCD axions\cite{axion}, and the correct
scale of neutrino masses\cite{mneut}. For a comprehensive review see
\cite{joethesis}. The scenario arises by considering generic quantum
corrections to the string action in KKLT style compactifications
\cite{KKLT} on a Calabi-Yau manifold with a ``Swiss-cheese'' style
geometry, and is immune to the usual fine-tuning problems of KKLT
compactifications. There is a danger that once we consider some
quantum corrections it is inconsistent to ignore all the others: this
is the Dine-Seiberg problem \cite{dineseiberg}. However, initial
studies suggest that the models are robust against further corrections \cite{loopjump, michele}. 

When compactified on a Calabi-Yau orientifold type IIB string theory
can be described at low energies by an effective $\mathcal{N}=1$ 4D
supergravity theory. In this paper we assume that this theory has the
matter content of the Minimal Supersymmetric Standard Model (MSSM). The Lagrangian is then determined by a K\"ahler potential $\mathcal{K}$, a superpotential $W$ and gauge kinetic functions $f_a$, which we may expand in terms of moduli fields $\Phi$, matter fields $C^{\alpha}$ and the two MSSM higgs fields $H_1$ and $H_2$ as follows:
\begin{equation} W=\hat{W}(\Phi)+\mu(\Phi)H_1 H_2 + \frac{1}{6}Y_{\alpha\beta\gamma}(\Phi)C^{\alpha}C^{\beta}C^{\gamma} + \dots, \end{equation}
\begin{equation} K=\hat{K}(\Phi,\bar{\Phi}) + \tilde{K}_{\alpha \bar{\beta}}(\Phi,\bar{\Phi})C^{\alpha}C^{\bar{\beta}} + \left( Z(\Phi,\bar{\Phi}) H_1 H_2 + h.c. \right) + \dots, \end{equation}
\begin{equation} f_a=f_a(\Phi).\end{equation}
We will review the origin of the MSSM soft terms but will gloss over the more technical aspects of the derivation which may be found in \cite{LVSsoftterms}. We turn first to the gauge kinetic functions, gauge couplings and gaugino masses.

\subsection{Gauge Couplings}
To be considered truly complete, any stringy model should have a sector which contains the Standard Model. We assume that such a sector can be found where the Standard Model will be supported on magnetised D7 branes. In order that the SM gauge groups not be too weakly coupled, these branes must wrap a small 4-cycle $\tau_s$ in the Calabi-Yau. The gauge kinetic functions may be computed by dimensional reduction of the DBI action whereupon one obtains
\begin{equation}
\label{eq:gkfns}
f_a = \frac{T_a}{4\pi} + h_a(F)S,
\end{equation}
where $S$ is the axio-dilaton field, $T_a$ is the K\"{a}hler modulus of the small cycle and $h_a$ is a topological function of the fluxes present on the D7 brane. These functions are in general dependent on the explicit brane configuration used to realise the Standard Model are and currently unknown for realistic scenarios outside of toroidal orientifolds \cite{torus} If the cycle size is increased, the fluxes become diluted and the gauge couplings become independent of the fluxes. In this dilute flux limit we may then write
\begin{equation}
\begin{array}{l}
f_{SU(3)} =\frac{T_s}{4\pi} , \\
 f_{SU(2)} =\frac{T_s}{4\pi}, \\
 f_{U(1)} =k_Y \frac{T_s}{4\pi}, 
\end{array}
\label{eq:diluteflux}
\end{equation}
where $k_Y$ is a generally model dependent normalisation for the $U(1)$ gauge field, which we regard as unknown. In the dilute flux limit this parameter will not affect the physics. 
 
 The gaugino masses are 
 \begin{equation} 
 M_a = \frac{1}{2} \frac{F^m \partial_m f_a}{ \mbox{Re} f_a},
 \end{equation}
 where $F^m$ are the supergravity moduli F-terms 
 \begin{equation}
 F^m = e^{\hat{K}/2}\hat{K}^{m\bar{n}}D_{\bar{n}}\bar{\hat{W}}
 \end{equation}
 which quantify the supersymmetry breaking. In the dilute flux limit this gives
 \begin{equation}
\label{eq:gaugino}
 M_1 = M_2 =M_3 = \frac{F^s}{2\tau_s} \equiv M,
 \end{equation}
where $M_i \equiv M_{SU(i)}$, so that the gaugino masses are universal at the compactification scale, which in these models is at the intermediate scale $m_s \ensuremath{\sim} 10^{11}\mbox{GeV}$.
 
\subsection{Soft Terms}
To derive the rest of the soft terms we use following standard expressions for the scalar masses, trilinear $A$-terms and the $B$-term \cite{brignole}:
\begin{eqnarray}
m_{\alpha}^2 & = & (m^2_{3/2} +V_0) - F^{\bar{m}}F^{n}\partial_{\bar{n}}\partial_{m}\log \tilde{K}_{\alpha} ,\\
A_{\alpha\beta\gamma} & =  & F^m \left[ \hat{K}_m \partial_m\log Y_{\alpha\beta\gamma} -\partial_m\log(\tilde{K}_{\alpha}\tilde{K}_{\beta} \tilde{K}_{\gamma}) \right], \\
B\hat{\mu} & = & (\tilde{K}_{H_1}\tilde{K}_{H_2})^{-1/2} \Bigg\{ e^{\hat{K}/2}\mu \left( F^m \left[ \hat{K}_m + \partial_m \log \mu - \partial \log (\tilde{K}_{H_1}\tilde{K}_{H_2})\right] - m_{3/2} \right) \nonumber \\
   && +(2m^2_{3/2}+V_0)Z - m_{3/2}\bar{F}^{\bar{m}}\partial_{\bar{m}}Z +m_{3/2}F^m\left[ \partial_m Z -Z \partial_m \log (\tilde{K}_{H_1}\tilde{K}_{H_2})\right] - \nonumber \\
    && \bar{F}^{\bar{m}}F^{n} \left[ \partial_m \partial_n Z - ( \partial_{\bar{m}}Z) \partial_n \log ( \tilde{K}_{H_1}\tilde{K}_{H_2})\right]\Bigg\} ,
 \end{eqnarray}
where $\tilde{K}_{\alpha\bar{\beta}} = \tilde{K}_{\alpha} \delta_{\alpha\bar{\beta}}$ ( without summing over $\alpha$) is the matter field metric. Using the metric for chiral matter fields derived in ref. \cite{kahler}
\begin{equation}
\tilde{K}_\alpha ~ \frac{\tau_s ^{\lambda}}{\mathcal{V}^{2/3}}k_\alpha (\phi)
\end{equation}
leads to the soft terms in the dilute flux limit \cite{LVSsoftterms}
\begin{equation}
\begin{array}{c}
M_\alpha = M   \\
m_\alpha =\sqrt{\lambda}M  \\
A_{\alpha\beta\gamma} = -3\lambda M   \\
B = -(\lambda+1)M  \\  
\end{array}
\label{eq:softwlambda}
\end{equation}
where $\lambda$ is the modular weight of the matter fields with
respect to the small cycles. In the minimal case where all branes are
wrapping the same cycle it was shown in ref. \cite{kahler} that
$\lambda=1/3$ so that eq.~\ref{eq:softwlambda} reduces to
\begin{equation}
\begin{array}{c}
M_\alpha = M   \\
m_\alpha = \frac{M}{\sqrt{3}} \\
A_{\alpha\beta\gamma} = -M   \\
B = -\frac{4M}{3}  \\  
 \end{array}
\label{eq:softterms}
\end{equation}
which interestingly reproduces the form of soft terms in the dilaton-dominated scenario of heterotic string models. It is this minimal case which we will consider in this paper.

It was demonstrated in \cite{LVSsoftterms} that it is impossible to
implement the $B$-term condition for $\mu>0$. The condition can be
satisfied for $\mu<0$ but only in a region of low $\tan\beta$ which
leads to a Higgs mass below the lower bound from LEP. We must then
assume that there exists a means to generate the correct $B$-term, for
example an NMSSM-style coupling $\alpha N H_1 H_2$, where $N$ is a
gauge invariant scalar which obtains a vev. There also might exist
vector-like matter between the string scale and the TeV scale which
could alter the RG equations and the low-scale soft terms. Perhaps the
most important \textit{caveat} is that we are using universal gaugino
masses (\ref{eq:gaugino}) at the intermediate scale. It is well known
that the standard model gauge couplings with MSSM field content
unify at the GUT scale. It is clear from eq. (\ref{eq:gkfns}) that
the required non-universality is provided in the LVS by the fluxes. We
can estimate the magnitude of the effect of the fluxes by running the
Standard Model $SU(2)$ and $SU(3)$ couplings to the intermediate scale
and noting the non-universality, obtaining  
\begin{equation}
 \frac{g_3^2}{g_2^2}\Bigg\vert_{m_s}  = \frac{M_3}{M_2}\Bigg\vert_{m_s} \approx 1.37.
\end{equation}
We adopt a compromise position of leaving the gaugino masses as universal at the string scale, but allowing the gauge couplings to differ by the amount above. In light of this discussion all results in this paper should be understood as being to leading order in the dilute flux approximation (\ref{eq:diluteflux}).

\section{Observables and the Likelihood}

\TABULAR[r]{|c|c|}{\hline
mSUGRA parameter & range \\ \hline
$m_0$ & 60 GeV to 1.5 TeV \\
$\tan \beta$ &  2 to 30 \\ \hline
SM parameter & constraint \\ \hline
$1/\alpha^{\overline{MS}}$ & 127.918$\pm$0.018 \\
$\alpha_s^{\overline{MS}}(M_Z)$ & 0.1172$\pm$0.002 \\
$m_b(m_b)^{\overline   MS}$ & 4.20$\pm$0.07 GeV \\
$m_t$ & 172.6$\pm$1.4  GeV\\ \hline
}{Input parameters \label{tab:inp}}

In calculating the likelihood we follow refs.\cite{Allanach:2005kz} and \cite{Allanach:2006jc}. We vary the six inputs shown in Table~\ref{tab:inp}. Several runs with varying starting points in parameter space were performed with an expanded parameter range $60<m_0<2000\mbox{GeV}$ and $2< \tan\beta <62$ which showed that consistent solutions of the RG equations only existed within the more limited region shown. To increase the efficiency of our simulations we restrict these parameters accordingly. The four SM inputs are constrained to lie within 4$\sigma$ of their central values. $1/\alpha^{\overline{MS}}$ and $m_b(m_b)$ are taken from ref.\cite{pdg}, $\alpha_s^{\overline{MS}}$ from ref.\cite{alphas} and $m_t$ from the most recent combined Tevatron analysis \cite{mt}. Given how small the experimental errors are we fix the muon decay constant $G_{\mu}$ to be $1.16637\times10^{-5}$ GeV$^{-2}$ and the mass of the Z vector boson to be $91.1876\mbox{GeV}$\cite{pdg}. To calculate the spectrum of the MSSM we use a modified version of \texttt{SOFTSUSY} 2.0.17 \cite{softsusy}. The $W$ mass and $\sin^2\theta_{\textup{eff}}$ are obtained with a code based on refs. \cite{Heinemeyer:2006px,Mwerror}.

\TABULAR{|cc|cc|cc|cc|}
{\hline
$m_{\chi_1^0}$ & 37  & $m_{\chi^\pm_1}$ & 67.7 & 
$m_{\tilde g}$ & 195 &
$m_{{\tilde \tau}_1}$ & 76 \\ $m_{{\tilde l}_R}$ & 88 & $m_{{\tilde t}_1}$ &
  86.4 &
$m_{{\tilde b}_1}$ & 91 & $m_{{\tilde q}_R}$ & 250 \\
$m_{{\tilde \nu}_{e,\mu}}$ & 43.1 & & & & & & \\
\hline}{Lower bounds applied to sparticle mass predictions (in GeV). \label{tab:bds}}

If a generated spectrum contains a sparticle whose mass lies below the 95\% lower bounds listed in Table~2 \cite{deAustri:2006pe} it is assigned zero likelihood. Similarly, if a point contains tachyonic sparticles or does not break electroweak symmetry in the correct way it is assigned zero likelihood. If a point has survived this far it is passed via the SUSY Les Houches Accord (SLHA) \cite{slha} to \texttt{micrOMEGAS} 1.3.6 \cite{micromegas} to calculate the dark matter relic density, the branching ratio of the rare decay $BR(B_s \rightarrow \mu^{+}\mu^{-})$ and the anomalous magnetic moment of the muon, $(g-2)_{\mu}$, and to the most up-to-date version of \texttt{SuperIso} \cite{superiso} to calculate the the isospin asymmetry of the decay $B\rightarrow K^* \gamma$ and the branching ratio $BR(b\rightarrow s\gamma)$.

\subsection{Observables}
To construct the likelihood we use a collection of observables from across electroweak physics, cosmology and B-physics.
\FIGURE{
\includegraphics[height=4.5cm]{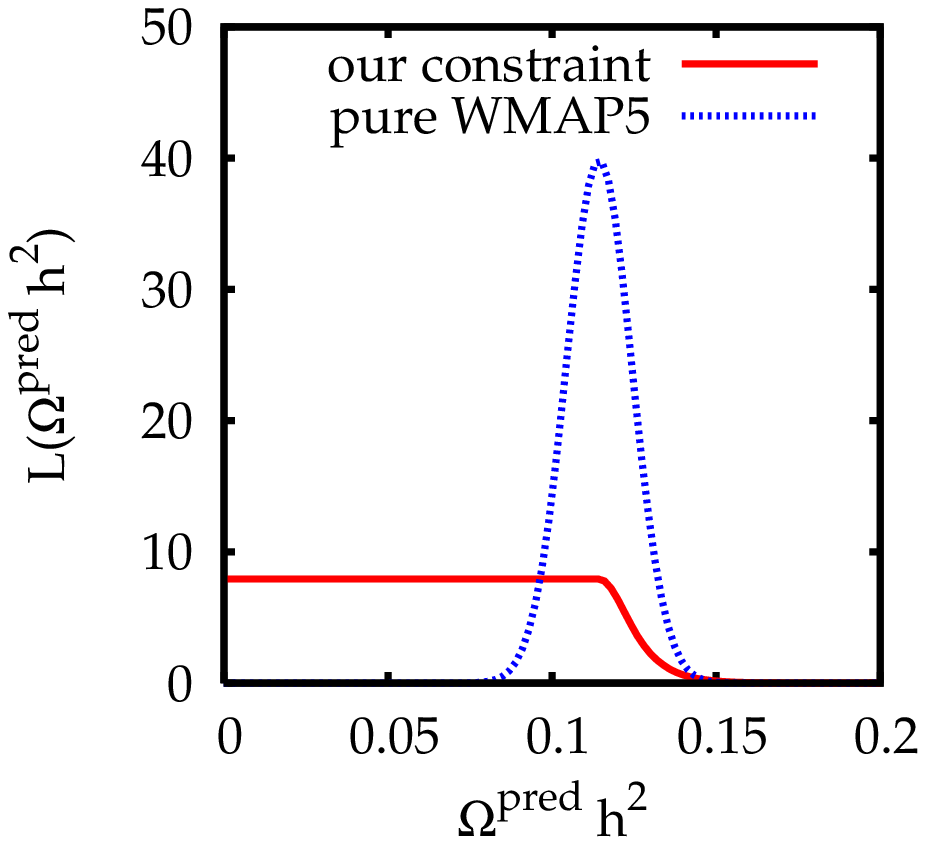}
\caption{Constraints for the dark matter relic density \label{fig:dmconstraint}}}

We use the newly released 5-year WMAP data \cite{WMAP} to constrain the cold dark matter relic density.
\begin{equation} \label{omegadm}
\Omega_{DM}h^{2} = 0.1143\pm0.0034
\end{equation}
As has been recently pointed out in ref.\cite{omegaconstraint} small
and currently undetectable changes in the expansion of the universe
before Big-Bang Nucleosynthesis (BBN) can lead to increases (but not
decreases) in the relic density by factors up to $10^4$. We must also
consider the possibility that the neutralino does not constitute the
entire relic density: there may be other components such as axions or
dark fluids. More recently, ref.\cite{RHNDM} have discussed the
effects of massive right-handed neutrinos on the mSUGRA relic
density. Contrary to common lore, the neutrino Yukawa couplings can
increase the relic density by more than an order of magnitude. We
therefore adopt the suggestion of \cite{omegaconstraint} and given a
prediction of the relic density $\omega$  we take the likelihood to be 
\begin{equation}
\frac{1}{c+ \sigma\sqrt{\frac{\pi }{2}}} \qquad (\Omega h^2 <0.1143),\qquad \frac{e^{ -(c - \omega)^2 / 2 \sigma^2 }}{c+\sigma\sqrt{\frac{\pi}{2}}}  \qquad (\Omega h^2 > 0.1143).
\end{equation}
In this way points below the WMAP bound are assigned an equal
likelihood, whereas those above are subject to a Gaussian likelihood
centred on $c=0.1143$ with standard deviation $\sigma = 0.02$, which
represents theoretical error in the prediction of the relic density. We
show this in Fig.~\ref{fig:dmconstraint}, along with the constraint we
would obtain from pure WMAP5.

The anomalous magnetic moment of the muon gains contributions from QED, hadronic vacuum polarisations and light-by-light processes. Evaluating the vacuum polarisation with $e^{+}e^{-}$ and $\tau$ data gives different results. Since up to date $e^+ e^-$ results are in line with earlier results, and those for the $\tau$ data are not, we restrict ourselves to the $e^+ e^-$ data and obtain \cite{SMmuon}
\begin{equation}
a_{\mu}^{SM} = (11659178.5\pm6.1)\times10^{-10}.
\end{equation}
When compared with the experimental result
\begin{equation}
a_{\mu}^{exp} = (11659208.0\pm6.3)\times 10^{-10}
\end{equation}
there exists a discrepancy
\begin{equation}
\delta \frac{(g-2)_{\mu}}{2} \equiv \delta a_{\mu} = a_{\mu}^{exp} - a_{\mu}^{SM} = (29.5\pm8.8)\times 10^{-10} 
\end{equation}
at the 3.4$\sigma$ level, which may be an indication of new physics. 
We evaluate this at one-loop with  \texttt{micrOMEGAS}, and then add in the logarithmic piece of the QED 2-loop calculation and the 2-loop stop-higgs and chargino-stop/bottom contributions \cite{2loopg-2, susymuon}.

To the experimentally measured mass of the W vector boson and the effective weak leptonic mixing angle \cite{wmass} we add the SM and MSSM theory errors detailed in \cite{Heinemeyer:2006px, Mwerror} to obtain
\begin{equation}
M_W = 80.398\pm 0.027 \mbox{ GeV}, \qquad \sin^{2}\theta^{l}_{w} = 0.23149\pm 0.00017.
\end{equation}
We utilise the full MSSM one-loop contribution, SUSY corrections of order $\mathcal{O}(\alpha\alpha_s)$ and $\mathcal{O}(\alpha_{t,b}^{2})$, as well as all relevant SM-like terms beyond one-loop order (see refs.\cite{Heinemeyer:2006px,Mwerror} for details).

We also use the LEP2 Standard Model Higgs mass bound that $m_h < 114.4\mbox{ GeV}$ at 95\% confidence level. We smear this by 3 GeV to represent the theoretical uncertainty in the \texttt{SOFTSUSY} prediction for $m_h$, as described in ref.\cite{Allanach:2006jc}.

Flavour-changing neutral currents are one of the areas most sensitive to new physics, particularly in the large $\tan\beta$ regime. With the accumulation of precision results from BABAR and Belle over recent years these observables have gained considerable discriminatory power.

The rare branching ratio $BR(b\rightarrow s\gamma)$ has been measured as \cite{bsgamma} $BR(b\to s\gamma) = (3.55\pm0.26)\times10^{-4}$.
We add in the SM and MSSM uncertainties as in ref.\cite{nazila0710} to obtain
\begin{equation}
BR(b\to s\gamma)=(3.55\pm0.72)\times 10^{-4}.
\end{equation}

For the branching ratio $BR(B_s\to \mu^{+}\mu^{-})$ the most recent upper bound from the Tevatron is \cite{bsmumu} 
\begin{equation}
BR(B_s \rightarrow \mu^{+}\mu^{-} )< 5.8\times10^{-8}
\end{equation}
at 95\% confidence level.

The current HFAG average of the branching ratio for the process $B_u \to\tau\nu$ is \cite{hfagbtnu}
\begin{equation}
BR^{exp}(B_u \rightarrow\tau\nu)=(1.41\pm0.43)\times10^{-4}.
\end{equation}
The Standard Model prediction of this branching ratio depends on
whether one determines the CKM matrix element $|V_{ub}|$ inclusive or
exclusive of semileptonic decays\cite{btnu1}. We
statistically average over these two values and get $
BR^{SM}=(1.12\pm0.25)\times 10^{-4}$, so that 
\begin{equation}
R^{exp}_{B\tau\nu} = \frac{BR^{exp}(B_u\to \tau\nu)}{BR^{SM}(B_u\to\tau\nu)} = 1.259\pm0.378.
\end{equation}
The MSSM contribution to this branching ratio is dominated by charged-Higgs contributions and to leading order in $\tan\beta$ gives
\begin{equation}
 R_{B\tau\nu}= \left[ 1-\left(\frac{m_B^2}{m_{H^{\pm}}^2}\right) \frac{\tan^2\beta}{(1+\epsilon_0 \tan\beta)} \right]^2,
\end{equation}
where $\epsilon_0$ is an effective coupling which takes into account the non-holomorphic correction to the down-type Yukawa coupling induced by gluino exchange \cite{isidori2006} and is given by
\begin{equation}
\label{eq:epsilon0}
\epsilon_0 = -\frac{2\alpha_s \mu}{3\pi M_{\tilde{g}}} H_2\left( \frac{M_{\tilde{q}_L}^2}{M_{\tilde{g}}^2} ,  \frac{M_{\tilde{d}_R}^2}{M_{\tilde{g}}^2} \right)
\end{equation}
where 
\begin{equation}
 H_2(x,y)= \frac{x\ln x}{(1-x)(x-y)} + \frac{y\ln y}{(1-y)(y-x)},
\end{equation}

The mass splitting of the $B_s$ meson has been measured by CDF to be \cite{deltamsexp}
\begin{equation}
\Delta^{exp} M_{B_s} = 17.77\pm0.12 \mbox{ps}^{-1},
\end{equation}
while the UTFit evaluation  of the standard model estimate is $\Delta^{SM}m_s = 20.9\pm2.6\mbox{ps}^{-1}$ \cite{UTFit}.
This gives us
\begin{equation}
R^{exp}_{\Delta M_{B_s}} = \frac{\Delta^{exp} M_{B_s}}{\Delta^{SM}M_{B_s}} = 0.85\pm0.12.
\end{equation}
The dominant MSSM contribution to $\Delta M_{B_s}$ comes from neutral
Higgs particles in double-penguin diagrams \cite{buras} and is given by
\begin{equation}
R_{\Delta M_{B_s}} = 1-m_b(m_b)m_s(m_b) \frac{64\pi \sin^2\theta_{\textup{eff}}}{\alpha_{\textup{em}} M^2_A S_0(m_t^2/m_W^2)} \frac{(\epsilon_Y \lambda_t ^2 \tan^2\beta)^2}{\left[ 1+(\epsilon_0 +\epsilon_Y\lambda_t^2)\tan\beta\right]^2 \left[ 1+\epsilon_0 \tan\beta\right]^2},
\end{equation}
where 
\begin{equation}
\epsilon_Y = -\frac{A_t}{16\pi^2\mu}H_2\left( \frac{M_{\tilde{q}_L}^2}{\mu^2} , \frac{M_{\tilde{u}_R}^2}{\mu^2} \right)
\end{equation}
and $\epsilon_0$ is the same as in eq.~(\ref{eq:epsilon0}) and $S_0$ is a Wilson coefficient which can be found in ref.\cite{S0}.
This does not take into account the charged Higgs and chargino box diagrams, however these provide only small contributions to the total splitting over the majority of parameter space.

The Standard Model predictions for the two above observables both
depend in some way on the mass of the bottom quark. Since this is also
something that we will be predicting and fitting to in our model, we
must consider the possible correlations between the Standard Model and
SUSY contributions. The main parameters which could depend on $m_b$ are
the B-meson mass $m_B$ and the mesonic decay constant $f_B$. We take
the $m_B$ from experiment to be 5.279 GeV\cite{pdg}, and so is not
correlated with any particular value of $m_b$. The decay constant
$f_B$ is calculated on the lattice and the associated value of $m_b$
is not an input but is derived by fixing a scale after which
predictions may be made. The value of $m_b$ obtained in this way is
consistent with all values that we will consider in this paper, see
ref.\cite{wingate}. 

Our final observable is the isospin asymmetry from the exclusive process $B\rightarrow K^* \gamma$, defined as
\begin{equation}
\Delta_{0-} = \frac{\Gamma(\bar{B}^{0} \rightarrow \bar{K}^{*0}\gamma) - \Gamma(B^-\rightarrow K^{*-}\gamma)} {\Gamma(\bar{B}^{0}\rightarrow\bar{K}^{*0}\gamma) + \Gamma(B^- \rightarrow K^{*-}\gamma)}.
\end{equation}
Data from BABAR \cite{isosymbabar} and Belle \cite{isosymbelle} have constrained this to be
\[ -0.018< \Delta_{0-} < 0.093 \]
at 95\% confidence level. We convert this to a Gaussian with central value and standard deviation
\begin{equation}
\Delta_{0-} = 0.0375 \pm 0.0289.
\end{equation}

\subsection{The Likelihood}
A prediction $p_i$ of one of the above observables (except $\Omega_{DM}h^2$, $BR(B_s\to \mu^{+}\mu^{-})$ and $m_h$ which we treat separately) is assigned the log likelihood
\begin{equation}
\ln \mathcal{L}_i = -\frac{(c_i - p_i)^2}{2 s_i^2} - \frac{1}{2}\ln(2\pi) - \ln(s_i)
\end{equation}
where $c_i$ is the measured central value and $s_i$ the standard deviation. The relic density likelihood is treated as in eq.~(\ref{omegadm}). The likelihood for the branching ratio of $B_s\to \mu^{+}\mu^{-}$ is calculated using the predicted value from \texttt{micrOMEGAS} and CDF Tevatron Run II data \cite{cslin}. For the Higgs we use a parametrisation of the search likelihood from LEP2, as used in ref.\cite{higgs}. We then calculate the combined likelihood for all observables
\begin{equation}
\ln\mathcal{L}^{tot} = \sum_i \ln\mathcal{L}_i,
\end{equation}
 equivalent to the assumption that our observables form a set of independent quantities.

\subsection{Priors}
In a Bayesian framework our ignorance is quantified through the posterior probability density function. Previously \cite{Allanach:2005kz,deAustri:2006pe}, this has been set to be 
\begin{equation}
\label{postpdf}
p(m_0, \tan\beta,s| \mbox{data}) = p(\mbox{data}| m_0, \tan\beta,s) \frac{p(m_0,\tan\beta,s)}{p(\mbox{data})}, 
\end{equation}
where $s$ are some Standard Model inputs and $p(\mbox{data}|m_0\tan\beta,s)$ is the likelihood. If we desire the posterior pdf for a particular parameter we marginalise over (i.e. integrate out) all other parameters. The natural measure which we use for marginalising comes from (\ref{postpdf}) so that if we wish to know the posterior pdf for $m_0$, for example, we calculate
\begin{equation}
\label{marginalised}
p(m_0|\mbox{data}) = \int \,d\tan\beta \,ds \quad p(m_0,\tan\beta,s|data).
\end{equation}
It was observed in \cite{naturalness} that this does not reflect the fact that $\tan\beta$ is a parameter derived from the more fundamental parameters $B$ and $\mu$. What we really desire is
\begin{eqnarray}
 p(m_0|\mbox{data}) =\int \,d\mu \,d B \,ds\quad p(m_0,\tan\beta,s|\mbox{data}) \delta(M_Z-M_Z^{exp}) \nonumber \\
= \int \,d\tan\beta \,d s \quad p(m_0,\tan\beta,s|\mbox{data})r(B,\mu,\tan\beta) \Big|_{M_Z=M_Z^{exp}},
\end{eqnarray}
where we have fixed  $M_Z$ at its experimental value since the experimental error is so small and $r(B,\mu,\tan\beta)$ is a Jacobian factor. To determine $r$ we follow ref.\cite{naturalness} and use the relations between $\tan\beta$, $M_Z$, $B$ and $\mu$ derived from the electroweak symmetry breaking conditions \cite{ewsb}
\[ \mu B = \frac{\sin 2\beta}{2} \left( \bar{m}^2_{H_1} + \bar{m}^2_{H_2} + 2\mu^2\right),\]
\begin{equation}
\mu^2 = \frac{\bar{m}^2_{H_1}-\bar{m}^2_{H_2}\tan^2\beta}{\tan^2\beta -1} - \frac{M_Z^2}{2} ,
\end{equation}
to obtain
\begin{equation}
r(B,\mu,\tan\beta)= M_Z \left| \frac{B}{\mu\tan\beta}\frac{\tan^2\beta-1}{\tan^2\beta+1} \right|,
\end{equation}
which will from now be referred to as the radiative electroweak symmetry breaking or REWSB prior. Viewed as a measure on parameter space, the REWSB prior assigns a higher weight to points with lower $\tan\beta$ and $\mu$.

\subsection{Convergence}
\FIGURE{
\includegraphics[height=7cm, angle=270]{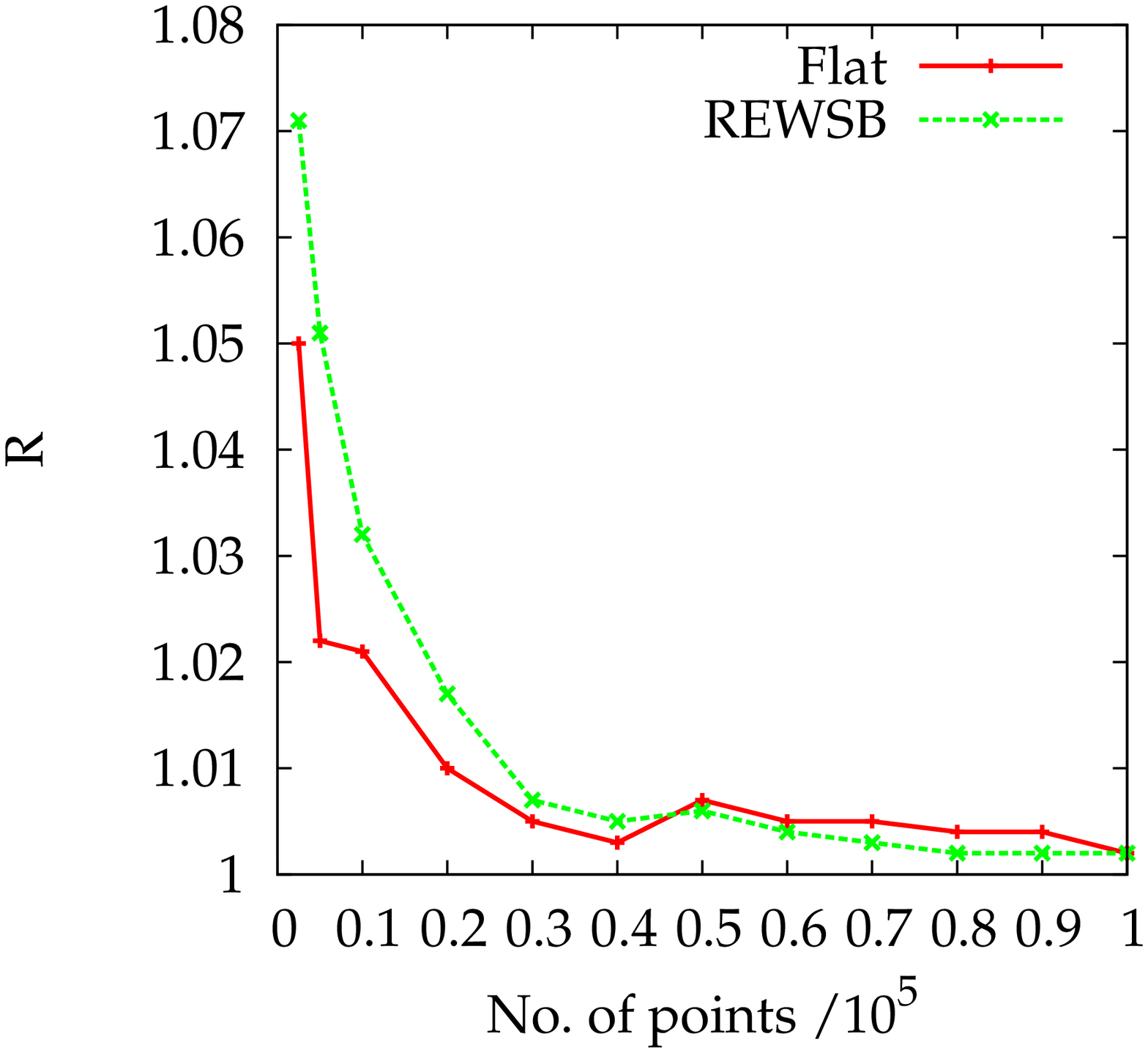}
\caption{No. of points plotted against R, the Gelman-Rubin convergence statistic. \label{fig:rvalues}}}
To accurately sample the posterior probability density we ran 10
independent MCMCs of length $10^5$ each. We discarded the initial 4000
steps as ``burn-in'' for the MCMCs. We use the Gelman-Rubin $\hat{R}$ statistic
\cite{Allanach:2005kz, gelmanrubin} to check for
convergence. In this test convergence is indicated by the value
$r<1.05$. With this measure all of our runs converge in less than
10,000 steps, and reach final R-values of 1.002 after $10^5$ steps.
We illustrate this in Fig.~\ref{fig:rvalues} by plotting the
Gelman-Rubin statistic against the number of steps taken. Convergence
is achieved here very rapidly, in around $10^4$ steps. This should be
contrasted with the mSUGRA case in ref.~\cite{Allanach:2005kz}, where
it took $6\times 10^5$ steps to converge. This is because the mSUGRA
likelihood distribution is a complicated multimodal distribution
spread out over a large region of parameter space, unlike the
relatively compact LVS likelihood distribution we will see below. The
overall efficiency of our simulation is 44\% (43\%) for $\mu>0$ and
23.9(22.8)\% for $\mu<0$ with flat (REWSB) priors
respectively. Throughout this paper when binning we use $75\times75$
bins, and all 2D plots are normalised to the maximum likelihood
bin. In all following 1D plots (except in Section 5) the vertical axis is the
posterior probability per bin.

\section{Likelihood Fits}
\label{sec:fits}

\FIGURE{ \mysixgraphs{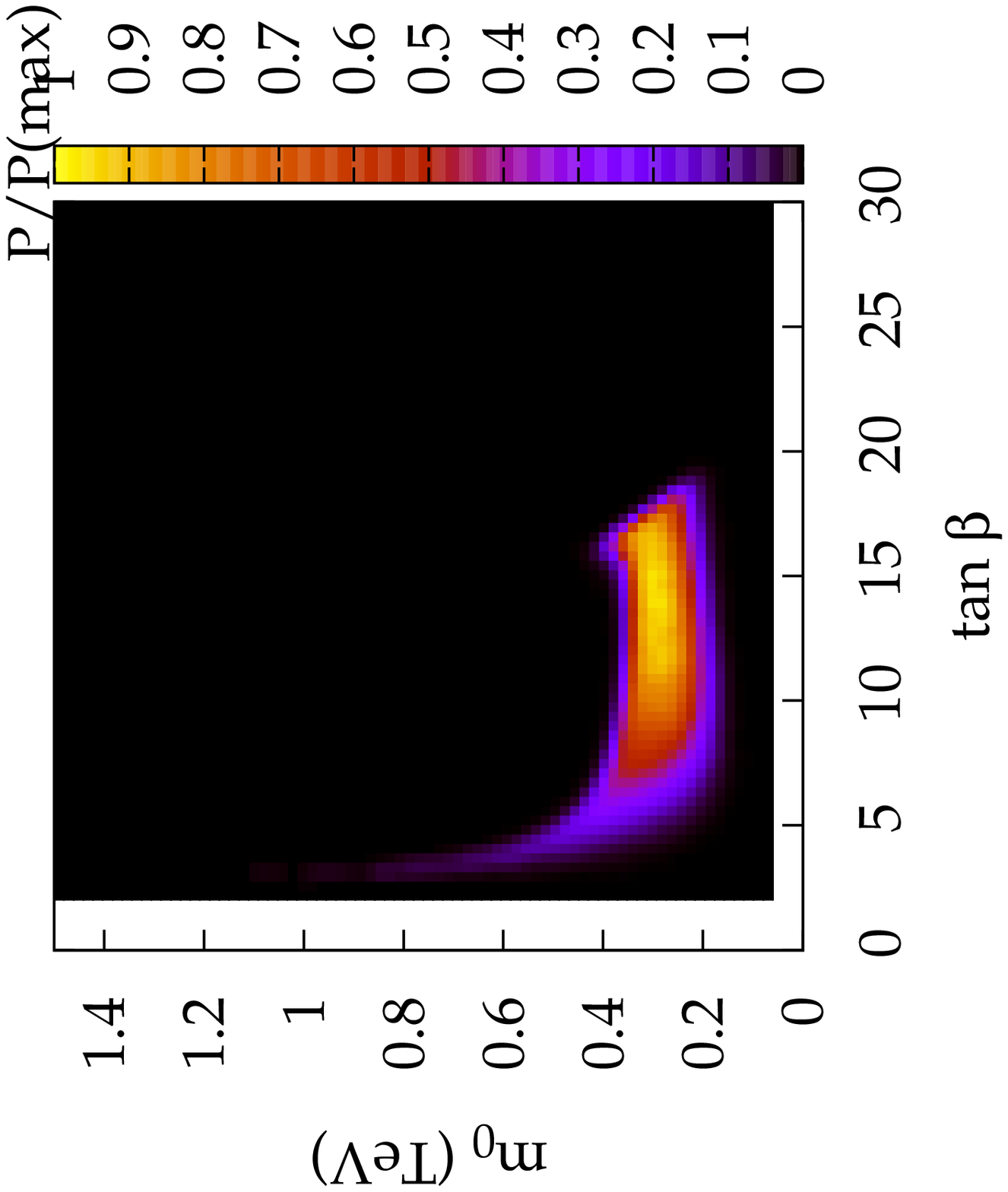}{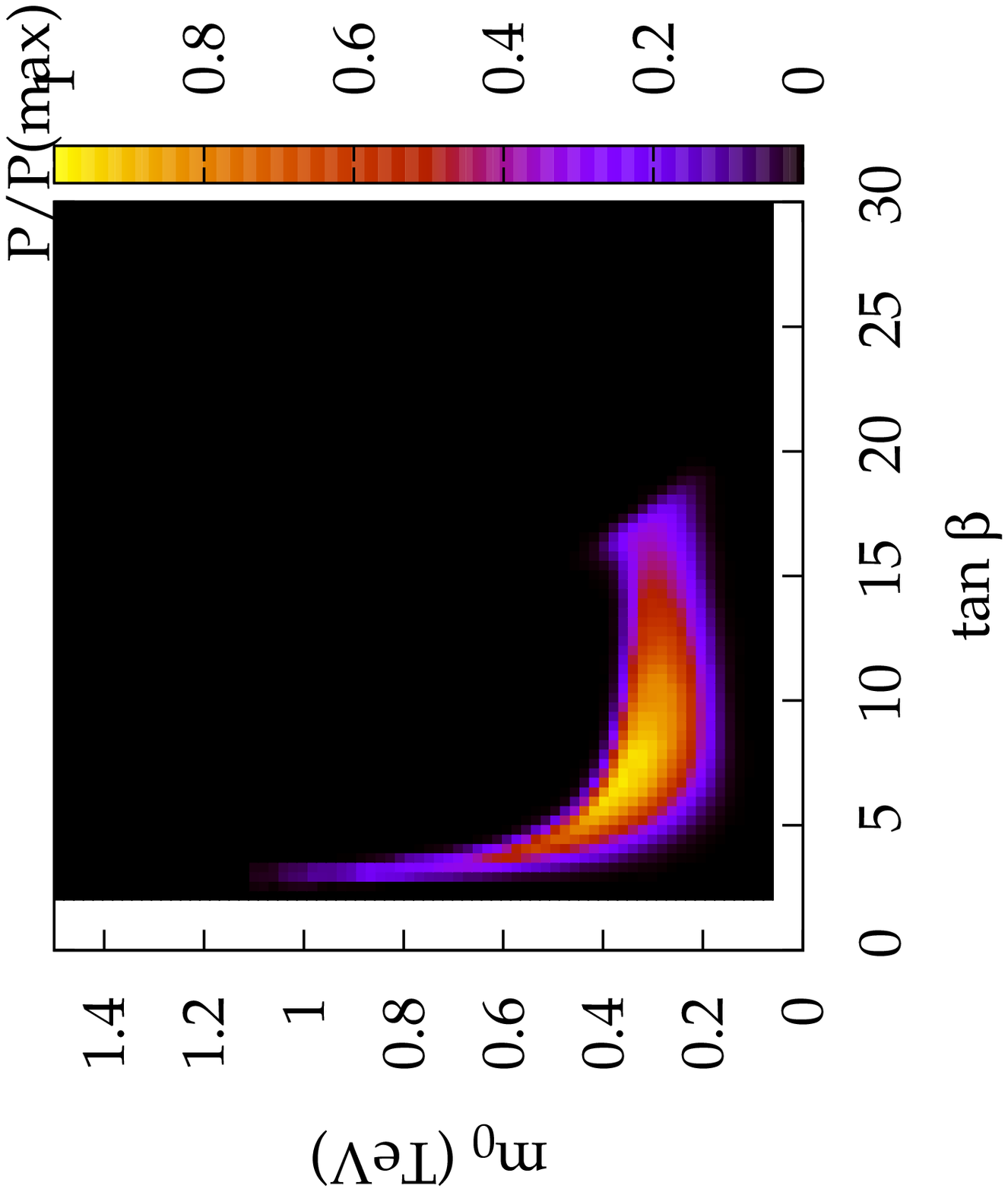}{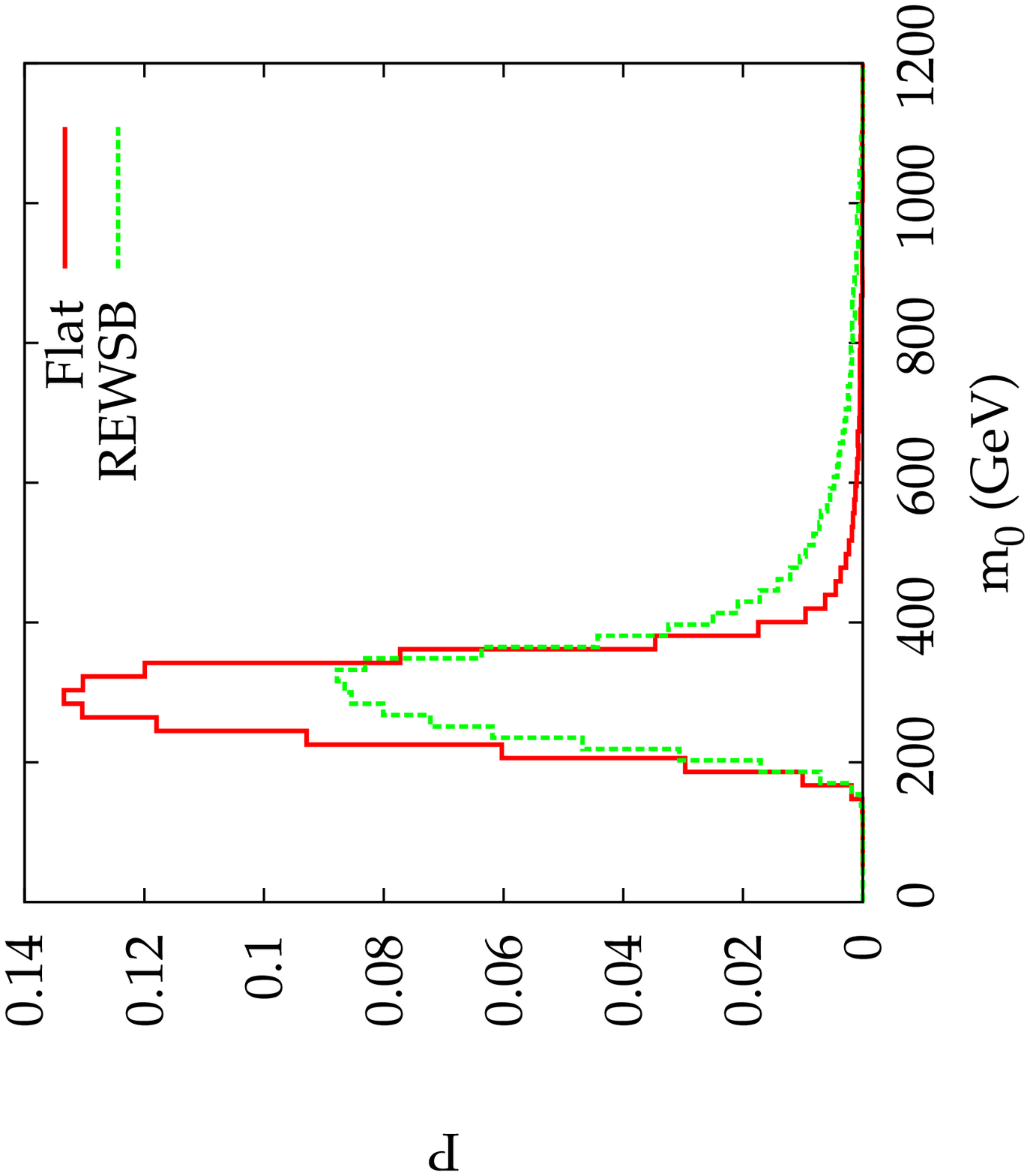}{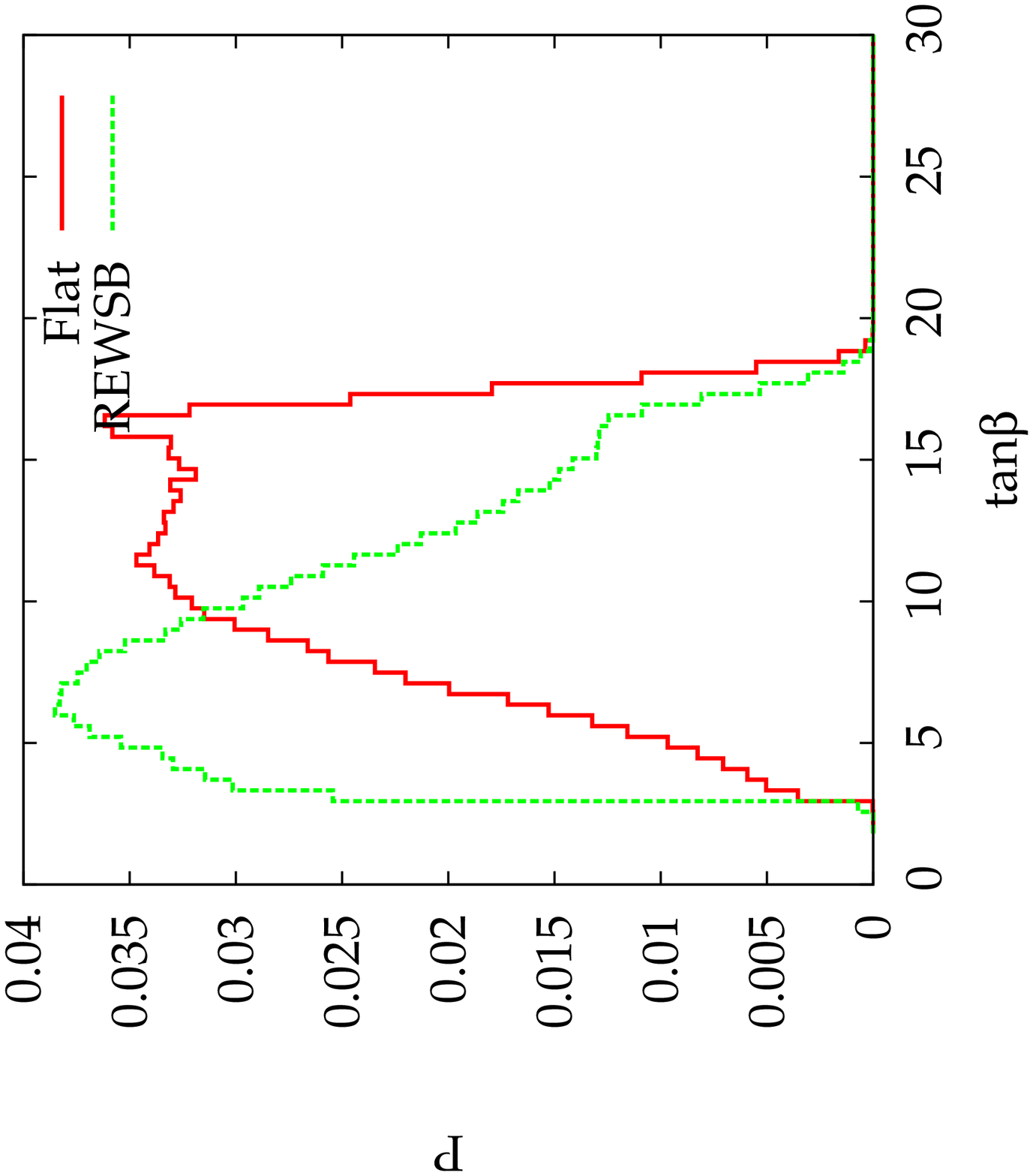}{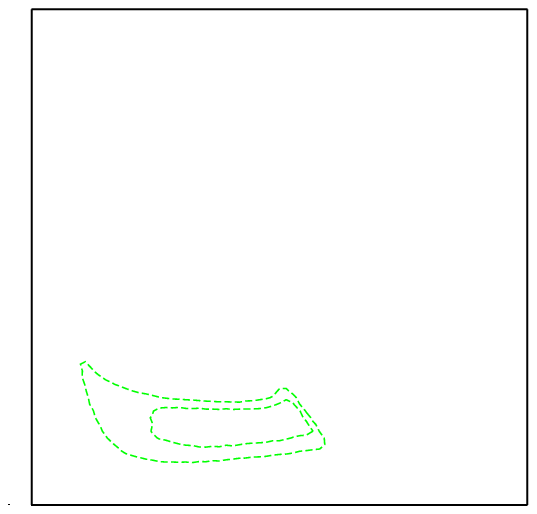}{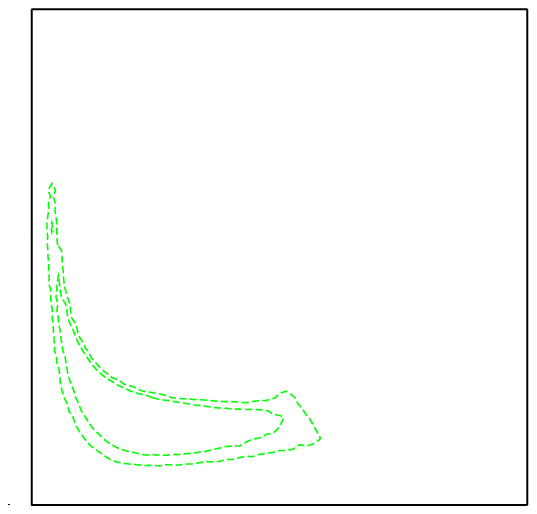} 
\caption{ Marginalised posterior pdfs in the $m_0$-$\tan\beta$ plane for $\mu>0$  and (a) flat priors and (b) REWSB priors with 68\% and 95\% c.l. contours shown, and the binned 1D distributions for (c) $m_0$ and (d) $\tan\beta$  also for $\mu>0$.\label{fig:m0tb}}}
Here we present the results of our analysis for values of $\mu > 0$,
with flat and REWSB priors. In Fig.~\ref{fig:m0tb}(a) and (b) we
present the posterior pdf marginalised to the $m_0$-$\tan\beta$
plane. We also show 68 and 95\% confidence limit contour lines. Since
$M_{1/2}$ and $A$ are determined linearly in terms of $m_0$ according
to eq.~(\ref{eq:softterms}), we do not present these pdfs. The sharp
cutoff at around $\tan\beta\approx 17$ is due to points to the right
of the cutoff having the stau as the LSP, which have been rejected
with zero likelihood. Fig.~\ref{fig:m0tb}(c) shows the 1D likelihood distribution of
the variable $m_0$. The majority of the likelihood is located between
200 and 400GeV for both sets of priors, with the upper bounds at 95\%
c.l. being $389.2\mbox{GeV}(616.2\mbox{GeV})$ for flat (REWSB)
priors. The REWSB priors have a fatter tail to higher $m_0$  due to
the boomerang shape of the posterior pdf, and the REWSB priors
favouring lower $\tan\beta$. Fig.~\ref{fig:m0tb}(d) shows the
likelihood distribution for $\tan\beta$. Solutions to all constraints
exist only in the range $2.5<\tan\beta<20$, and the 95\% upper bounds
are $17.1(16.0)$ for flat (REWSB) priors.  Although $\tan\beta$ is
difficult to determine experimentally, if measurements showed that we
live in a high $\tan\beta$ region (as favoured by mSUGRA for example
\cite{Allanach:2005kz}) this would discriminate against the Large
Volume Scenario. 

\TABULAR{|c|ccc|c|ccc|}{\hline
 & Flat  & REWSB  & $\Delta \chi^2$ & & Flat & REWSB & $\Delta\chi^2$\\ \hline
$m_0$/GeV & 300.7  & 394.7 & - &    $BR(b\to s\gamma)$ & 1.43 & 0.29 & 0.11  \\
$\tan \beta$ &  14.7 & 6.6 & - &    $BR(B_s\to\mu^+ \mu^-)$ & 0 & 0 &0\\ 
$\Omega_{DM} h^2$ & 0 & 0 & 0 &     $\sin^2 \theta_{\textup{eff}}$ & 0.02 & 0.04 & 0.01\\
$M_h$ & -0.48 & -0.35 & 0.10 &       $M_W$ & 0.94 & 1.12 & 0.04\\
$(g-2)_{\mu}$ & 5.48 & 9.54 & 0.20&  $\Delta_{0-}$ & 3.55  & 2.31 & 0.16\\ 
$\Delta M_{B_s}$ & 1.56 & 1.56 & 0& $BR(B_u \to\tau\nu)$& 0.57 & 0.48 & 0.01 \\ \hline
  &&&&                               $\chi^2$ (total) & 13.6 & 15.3 & 0.83\\ \hline
}{Best fit points for $\mu>0$ and statistical pull of observables. $m_0$ and $\tan\beta$ are the parameter space values we generate. All other numbers are $\chi^2$ values. The third column gives the statistical error on our estimates of the $\chi^2$ values for the flat prior best-fit point, as described in the text. \label{tab:bestfit}}

In Table \ref{tab:bestfit} we show the details of the best fit points for the flat and REWSB priors. For $m_0$ and $\tan\beta$ we show the values of these parameters, and for all other quantities we give the $\chi^2$ that quantity makes to the total $\chi^2$ of the fit. The third column presents our estimate of the uncertainty we have in our estimate of the absolute minimum of the minimum $\chi^2$ of all examples for the case of flat priors. To compute this we treat the best-fit point of each of the ten chains as an independent estimation of the minimum $\chi^2$ for the observables. From these ten points we  calculate the standard deviation of the minimum $\chi^2$ value, which we take as the uncertainty in our estimates. Given an infinite run time we would expect the flat and REWSB priors to converge to the same best-fit point. The fact that these are different is merely indicative of the finite length of our Markov chains. The neutralino makes up around two-thirds of the dark matter relic density and the lightest CP-even Higgs mass is in both cases around
$115\mbox{ GeV}$, just above the lower bound set by LEP2. Both sets of
priors slightly underpredict $M_W$ by about $1\sigma$. $\Delta
M_{B_s}$ varies by less than one percent over the entire parameter
space, and $R_{\Delta M_{B_s}}$ is never less than 0.99. Similarly, the
branching ratio $BR(B\to\tau\nu)$ does not exhibit much variation, and
$R_{B\tau\nu}$ is always above 0.87, inside the $1\sigma$ bounds we derived in Section 2. This is in
agreement with what was found in ref.\cite{susyspace}: that these two observables do not impose large changes to the fitted parameters.
The flat priors have an overall slightly smaller $\chi^2$ value, indicating a better fit to all observables, and so for the remainder of this paper we take this to be the ``true'' best-fit point. The REWSB priors have an overall better fit to the B-observables: a $\chi^2$ value of 4.6 versus 7.1 for the flat priors.  However, due to the value of $\tan\beta$ in the REWSB case being just under half that of the flat case, the supersymmetric contributions to the anomalous magnetic moment of the muon are very small for REWSB, which leads to the best-fit point of the flat priors having a smaller overall $\chi^2$ value. The reason that the relic density has no $\chi^2$ associated with it is because the value of the best-fit point falls below the WMAP central value, and does therefore not incur a likelihood penalty as discussed in Section 3.1.

We may perform a simple hypothesis test by computing the P-value
associated with the $\chi^2$ of the best-fit point. To do this
requires that we know the number of statistical degrees of freedom of our
fits. We argue that this is eight: we have fourteen observables
defined above (if one includes $m_t$, $m_b$, $\alpha_s$ and
$\alpha^{-1}$) and six model parameters: $m_0$, $\tan\beta$ and the
four SM observables just mentioned which are varied in our MCMC runs.
The P-value of the best-fit point is then $P_{flat}=0.093$. Taking a standard significance level of 0.05, we may then say that the LVS is indeed consistent with the available data.

\subsection{Dark matter}
\FIGURE{ \onegraph{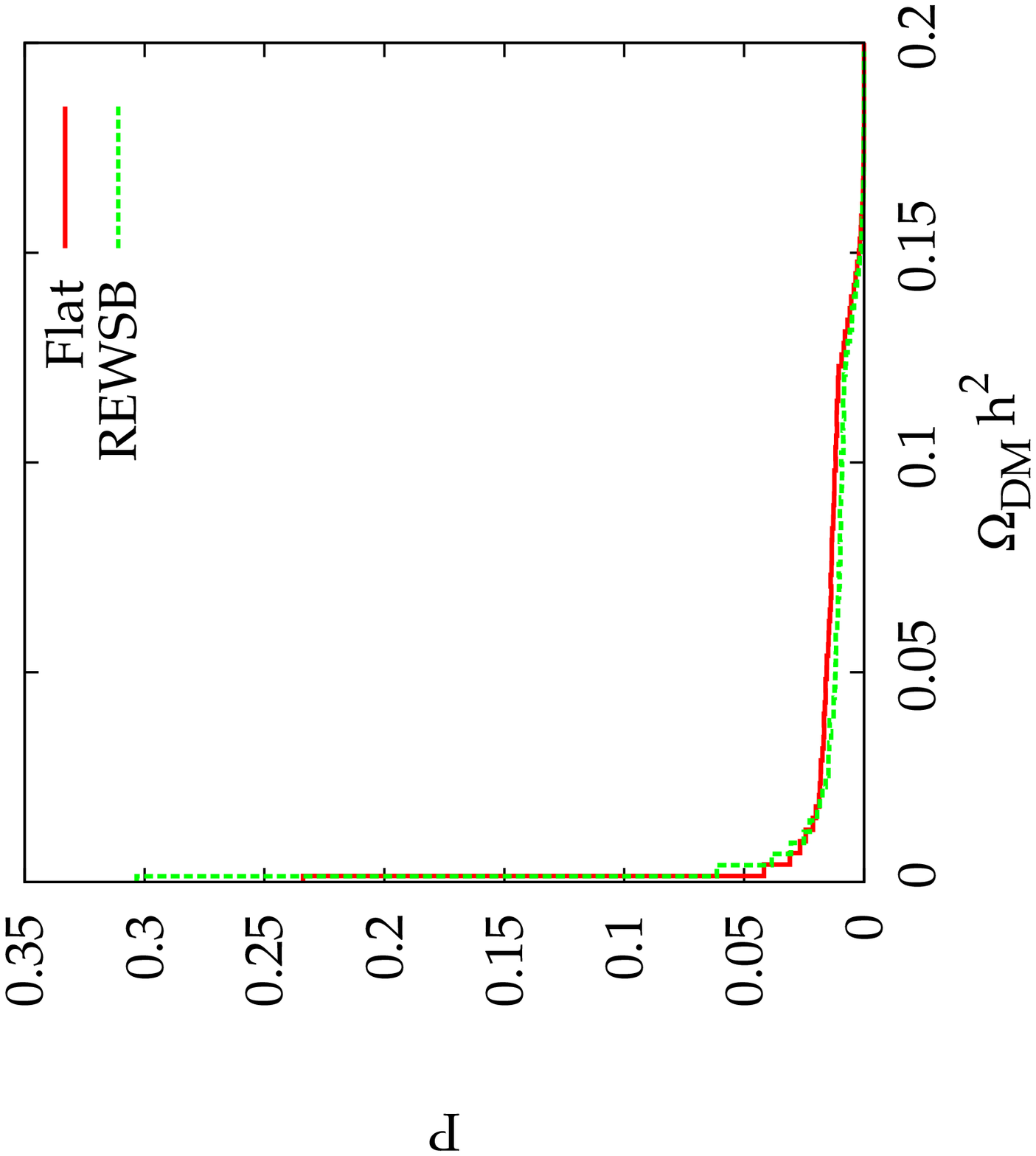} \caption{1-D posterior distribution for the relic density for flat and REWSB priors for $\mu>0$. \label{fig:dm}}}
In Fig.~\ref{fig:dm}  we show the 1D likelihood distribution for the dark matter relic density. The most striking feature for both sets of priors is the large spike in likelihood near the origin. To elucidate what is responsible for this we turn to discussing channels of relic density depletion.

We would like to assign probabilities to the possible relic density
depletion processes. To this end we follow ref.\cite{Allanach:2005kz}:
the stau co-annihilation region is where $m_{\chi_1^0}$ lies within
10\% of $m_{\tilde{\tau}_1}$, the $h^0 / A^0/Z$ pole region is where
$2m_{\chi_1^0}$ is within ten percent of $m_{h^0} / m_{A^0}/Z$,
respectively,  and the stop co-annihilation is situated where $m_1^0$
is within 30\% of $m_{\tilde{t}_1}$, since this channel is
particularly efficient. Since the lightest neutralino is always more
massive than $120\mbox{ GeV}$ the $h^0$-pole and Z-pole regions are
completely inaccessible throughout the parameter space. Similarly,
there is a negligible amount of stop co-annihilation. However for both
sets of priors the probability is higher than 99.5\% that we are in
both stau co-annihilation and A-pole regions. This is in sharp
contrast with mSUGRA, where the $A^0$ pole and stau co-annihilation
regions do not overlap and where the $A^0$ pole region is only found
at high $\tan\beta$.

To further investigate the spike we filter our chain of points, keeping only those with $\Omega_{DM}h^2 < 5\times 10^{-3}$. To obtain such a small relic density requires (at least) one of the annihilation channels to become extremely efficient, which should occur when the masses of some of the neutralino and some of the particles above become degenerate. We therefore plot in Fig.~\ref{fig:chistau}(a) the mass differences $m_{\tilde{\tau}}-m_{\chi^0_1}$ and $m_A - 2m_{\chi_1^0}$ for the filtered spike region, and for both sets of priors.
\TABULAR{|c|c|c|}{ \hline 
 & Flat & REWSB \\ \hline
1. & $\chi\chi\to b\bar{b}$ (55\%) & $\chi\chi\to t \bar{t}$ (50\%) \\
2. & $\chi\chi\to\tau\bar{\tau}$ (9\%) & $\chi\chi \to b\bar{b}$ (27\%)\\
3. & $\chi\tilde{\tau}\to\gamma\tau$ (7\%) & $\chi\chi\to\tau\bar{\tau}$ (5\%) \\
4. & $\tilde{\tau}\tilde{\tau}\to\tau\tau$ (7\%) & $\chi\tilde{\tau}\to\gamma\tau$ (2\%) \\
5. & $\chi\chi\to\ t \bar{t}$ (4\%) & $\chi\tilde{e}\to\gamma e$ (2\%) \\ \hline
}{ Top five relic depletion channels for the best-fit points in Table \ref{tab:bestfit} for $\mu>0$. \label{tab:dmchannels}}

\FIGURE{
\twographs{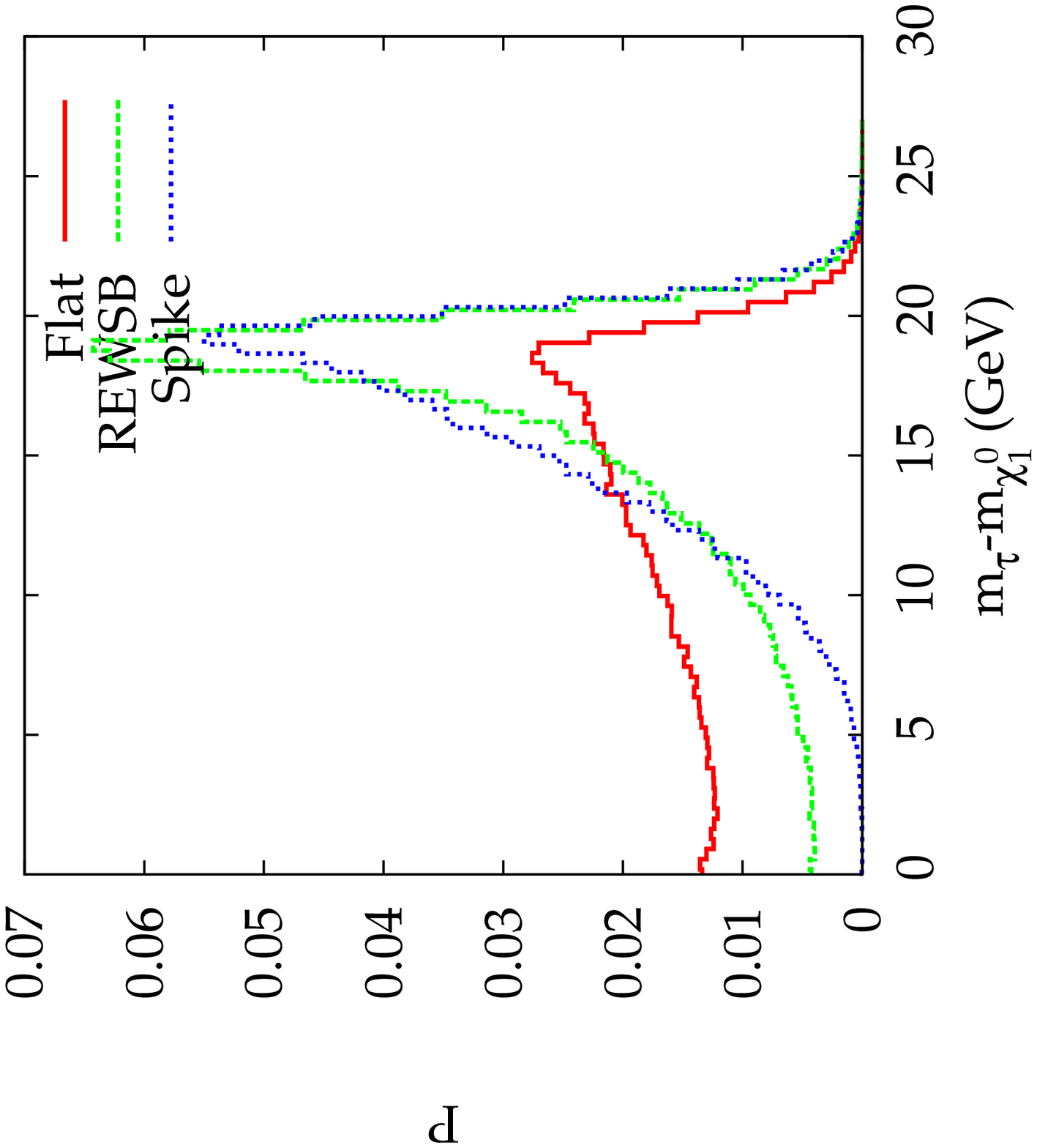}{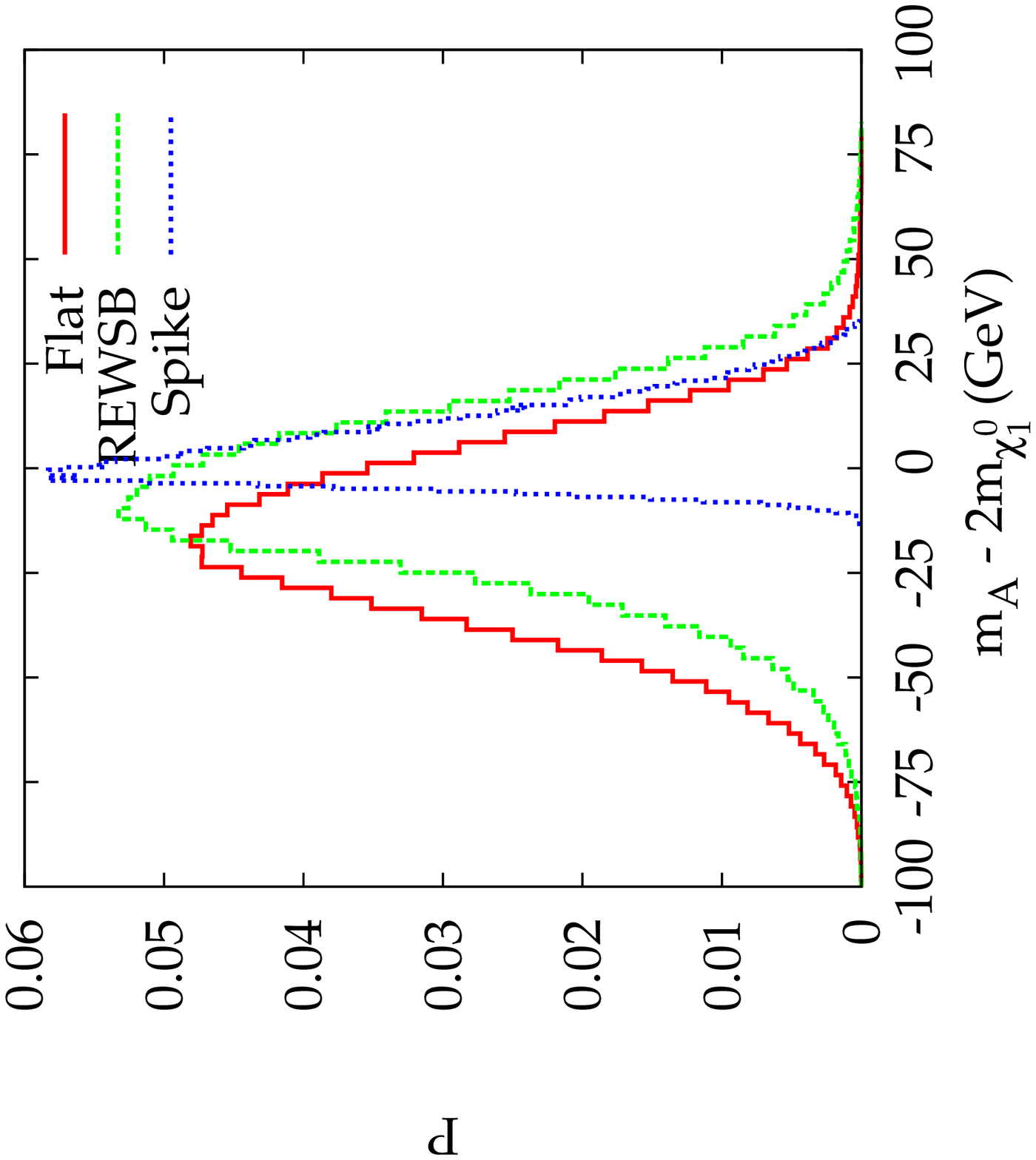}
\caption{Pdfs for the mass splitting between (a) $\chi_1^0$ and $\tilde{\tau}$ and (b) between $m_A$ and $2m_{\chi_1^0}$ with flat and REWSB priors, and for the filtered spike region discussed in the text for $\mu>0$. \label{fig:chistau}}}

The stau-neutralino mass splitting is always less than $25\mbox{ GeV}$, peaking at just less than $20\mbox{ GeV}$. This peak is more pronounced with the REWSB prior and is associated with the spike region. However, the spike pdf does not extend down to the degenerate region of the plot and stau co-annihilation does not significantly contribute to the relic density depletion in the spike.

In Fig.~\ref{fig:chistau}(b) we show the distribution of the mass difference $m_A - 2m_{\chi_1^0}$. The maximum of the plot occurs near $-20\mbox{ GeV}$ for the flat and REWSB priors, and for the filtered spike region the peak is nearly exactly at zero. The implication of this is that the spike is associated with a region where $m_A \approx 2m_{\chi^0_1}$. There still remains the question as to whether the spike is a region of high likelihood, or a large region of average or low likelihood. As we shall show in section \ref{sect:profile}, it is the latter that is the answer, and that the spike is an example of a so-called volume effect.
\FIGURE{\onegraph{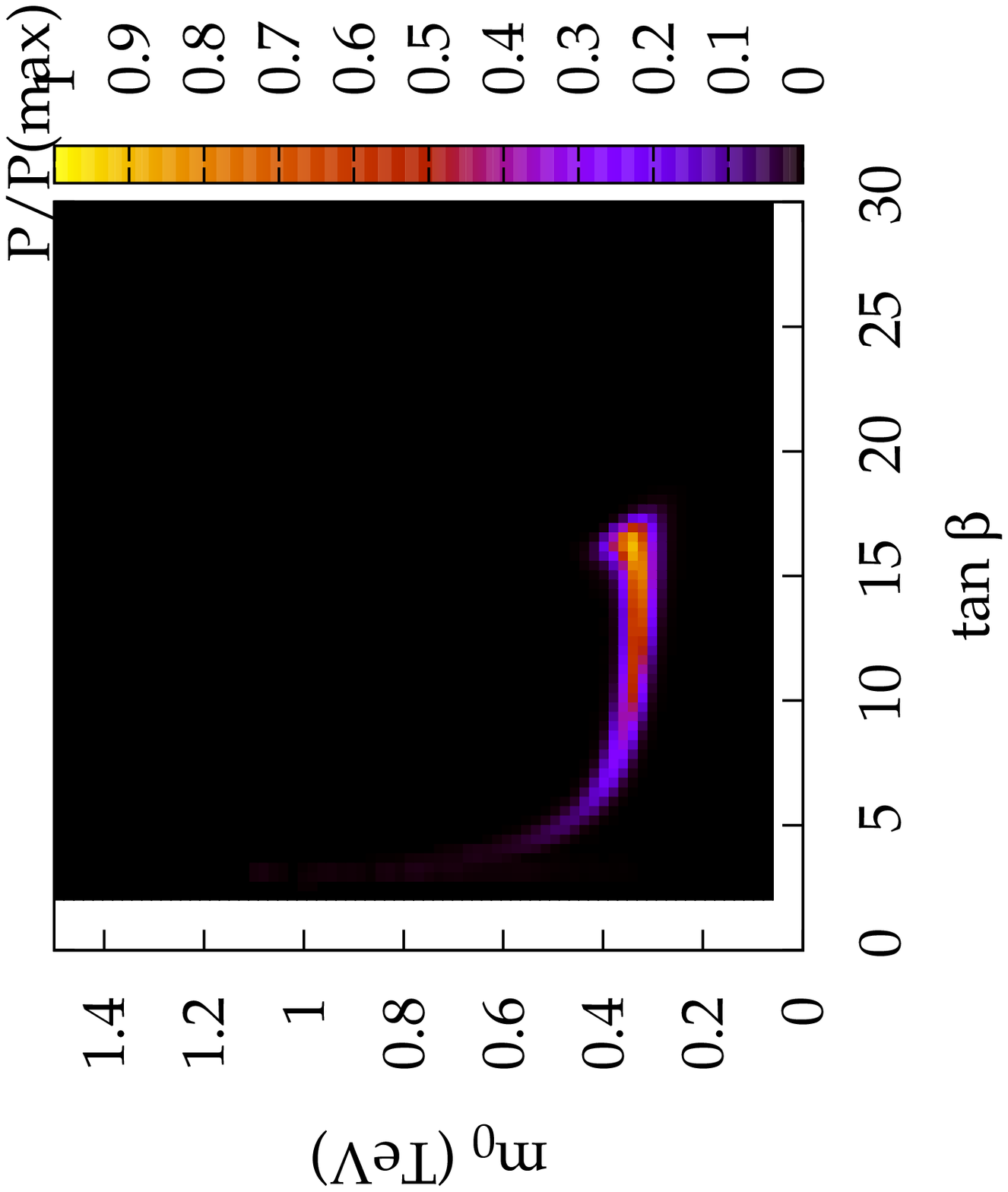}\caption{Posterior pdf marginalised to the $m_0$-$\tan\beta$ plane with WMAP5 dark matter constraint and $\mu>0$. \label{WMAPm0tb}}}

Table \ref{tab:dmchannels} shows the top 5 relic depletion channels
for the best fit points above, as calculated by
\texttt{micrOMEGAS}. The channels $\chi\chi\to b\bar{b}$, $t\bar{t}$
and $\tau\bar{\tau}$ are all s-channel interactions mediated by the
$A^0$ Higgs. These channels are responsible for 68\%(82\%) of the
relic density depletion for the flat (REWSB) best-fit points. This is
in agreement with what we would expect from Fig.~\ref{fig:chistau}.
Subdominant processes include stau co-annihilation, selectron
co-annihilation and $\tilde{\tau}\tilde{\tau}\to\tau\tau$.

Although we have purposefully left open the door open for exotic dark
matter in our choice of relic density constraint, the scenario that is both minimal and most
discussed in the literature is where the relic density is
entirely neutralino. To allow comparison with this case we reweight
our Markov chains so that likelihood function is a symmetric Gaussian
centred on 0.1143 with standard deviation 0.02, which is the pure
WMAP5 constraint shown in Fig.~\ref{fig:dmconstraint}, and then
re-analyse all our data.   Fig.~\ref{WMAPm0tb} shows the resulting
likelihood distribution in the $m_0$-$\tan\beta$ plane. As expected,
the distribution is much more constrained when compared with
Fig.~\ref{fig:m0tb}, with most of the loss coming from the low $m_0$
region. Points in the low $m_0$ region have increased dark matter
annihilation cross-sections due to lower sparticle masses and so the
relic density tends to be low, and it was this area which was
responsible for the spike in Fig.~\ref{fig:dm}.

\FIGURE{
\sevengraphs{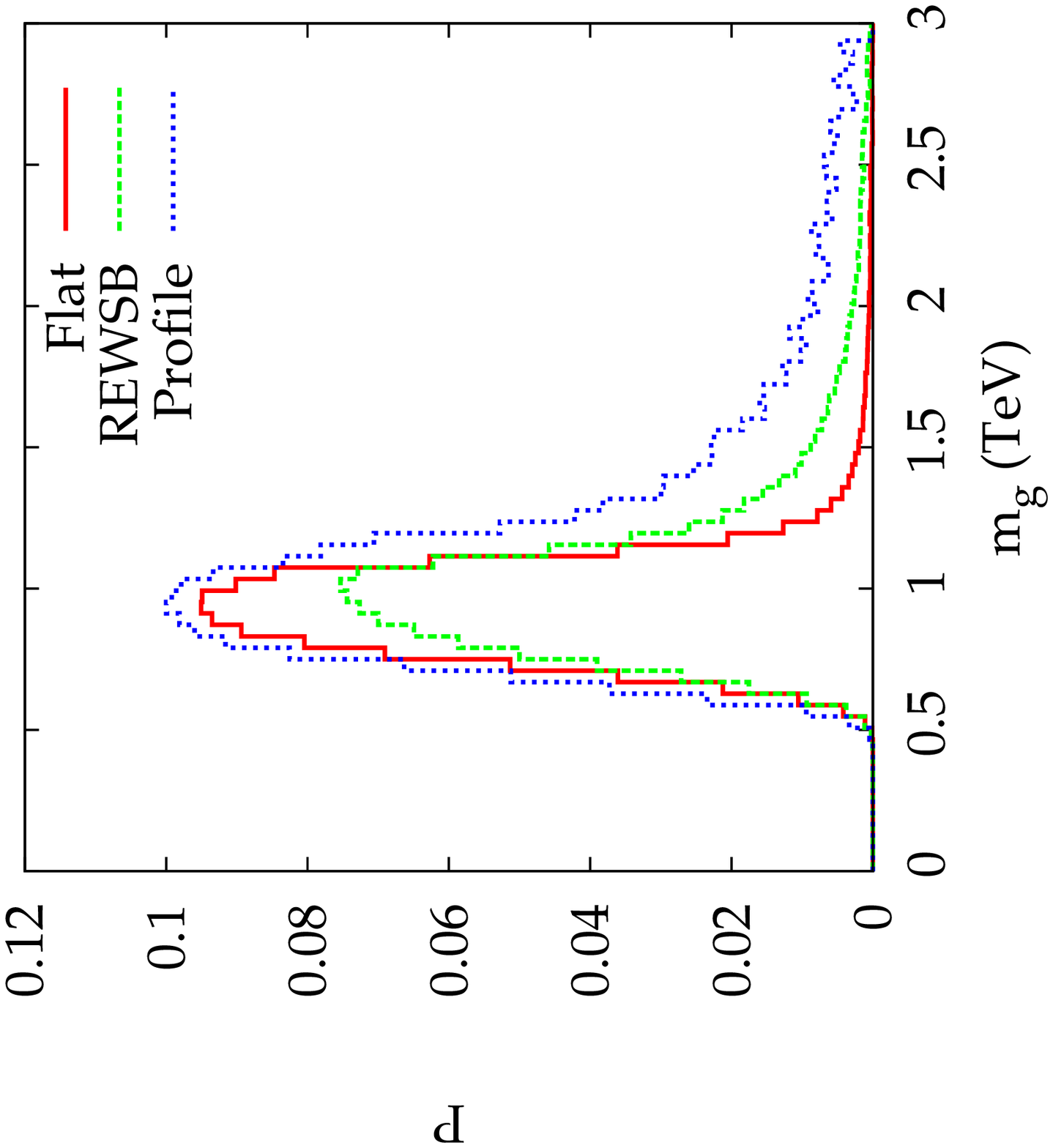}{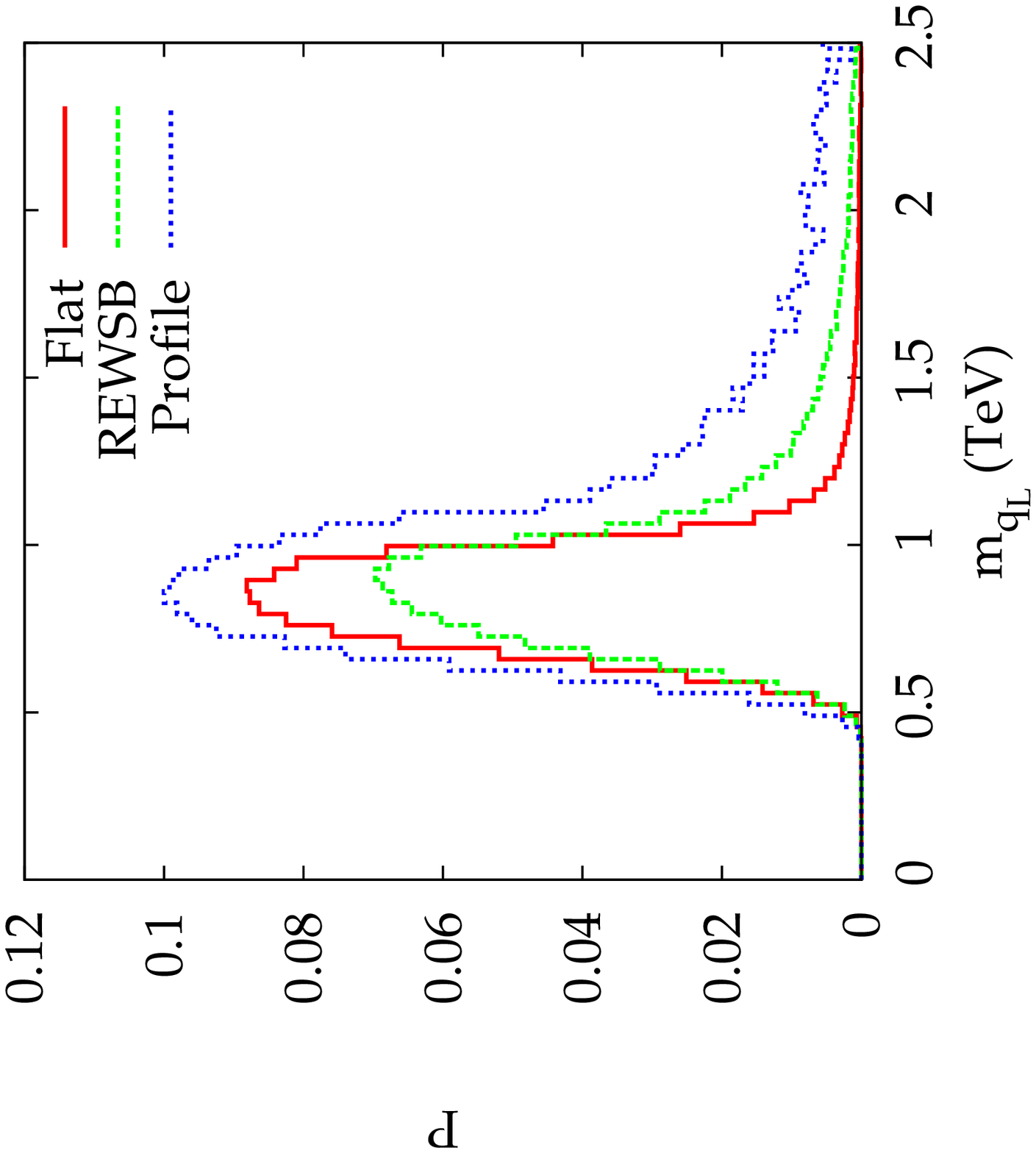}{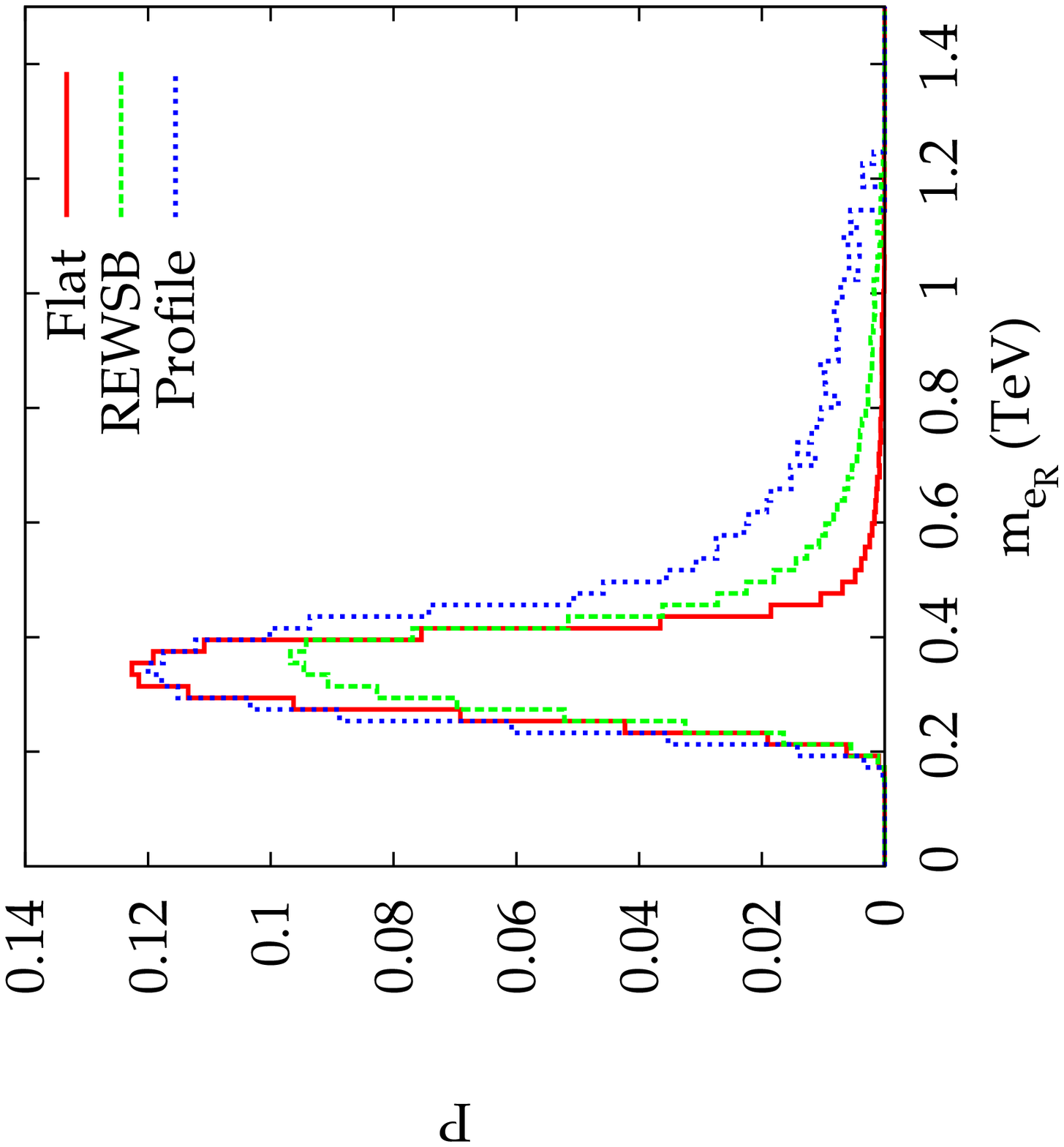}{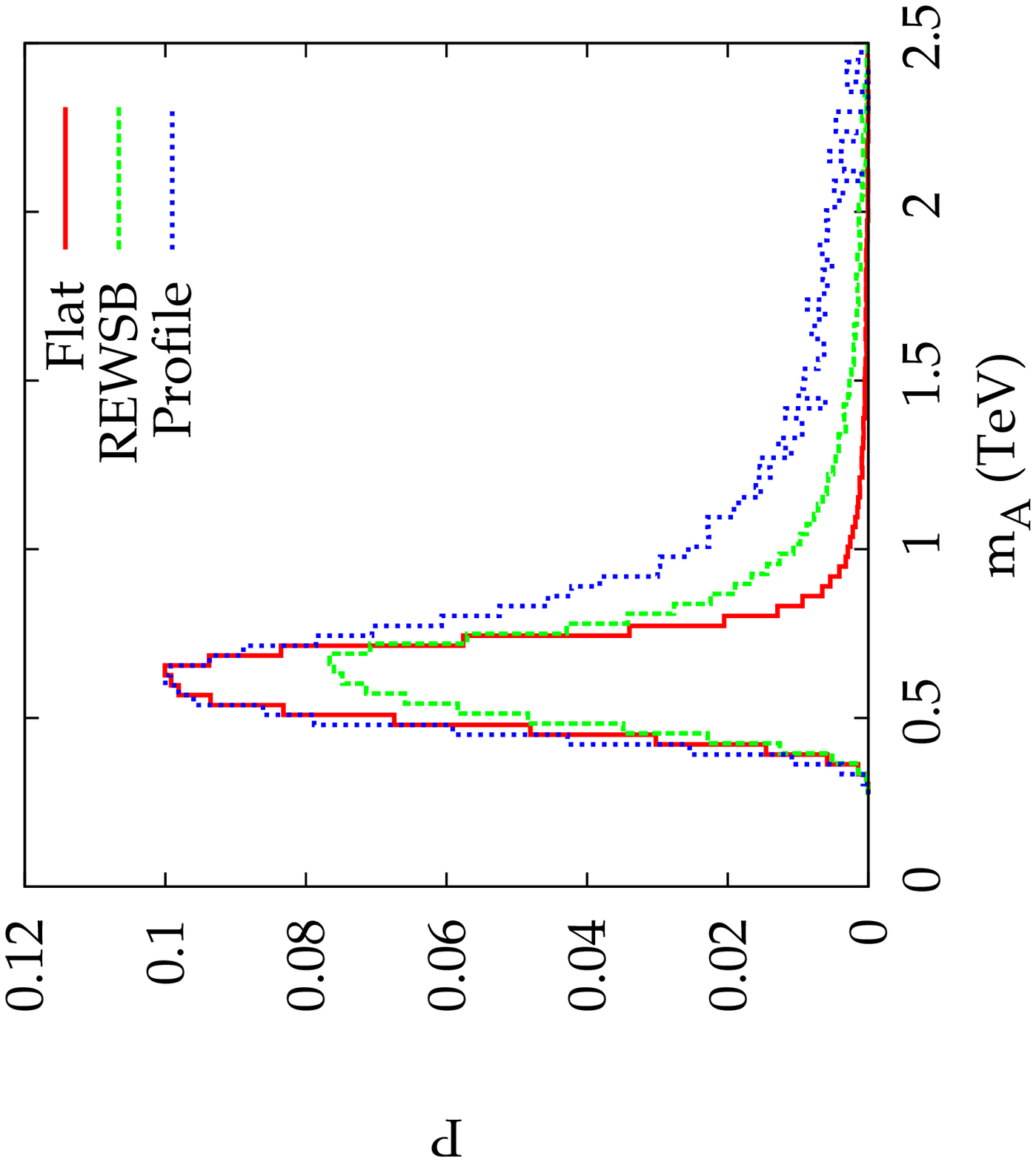}{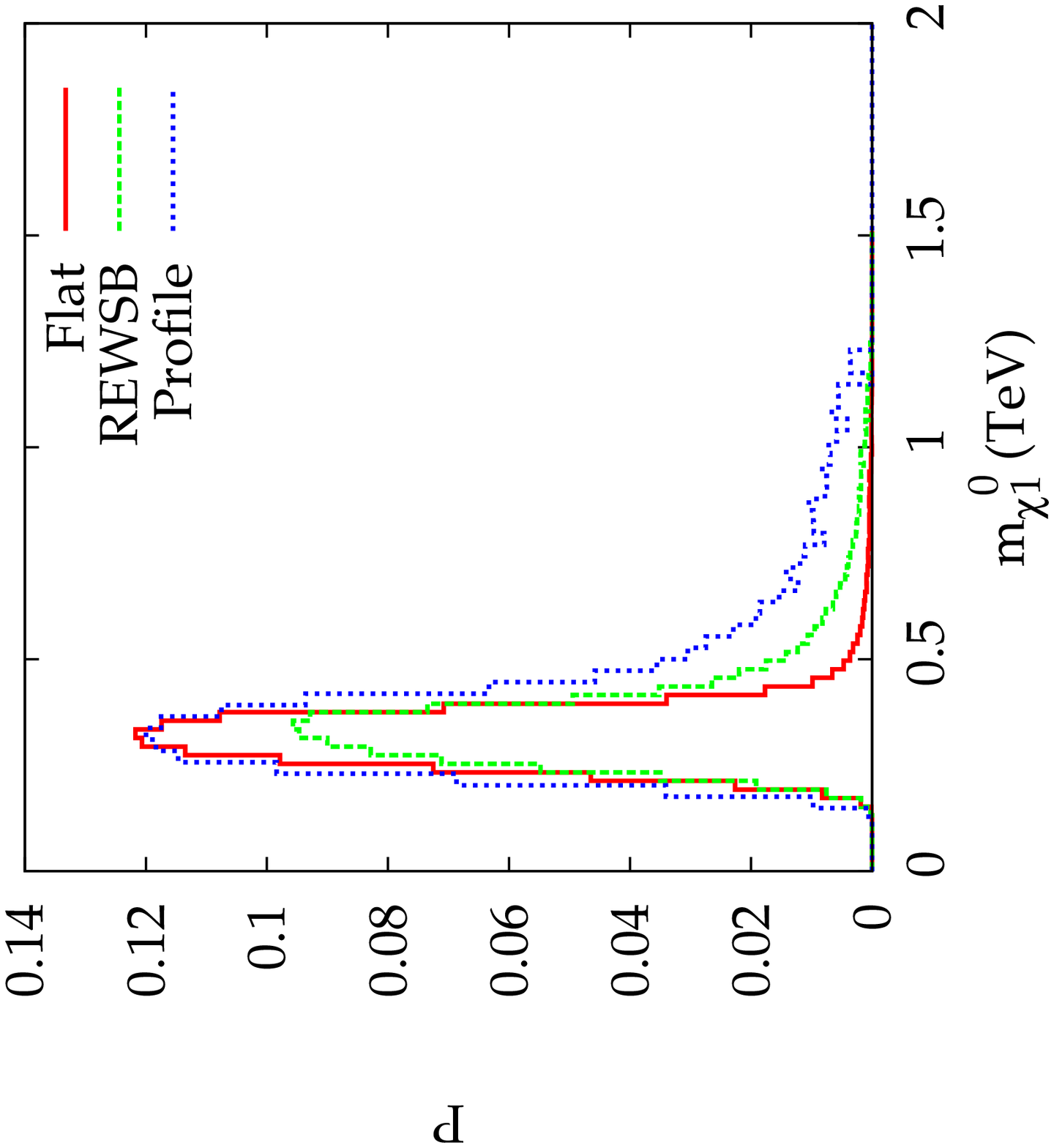}{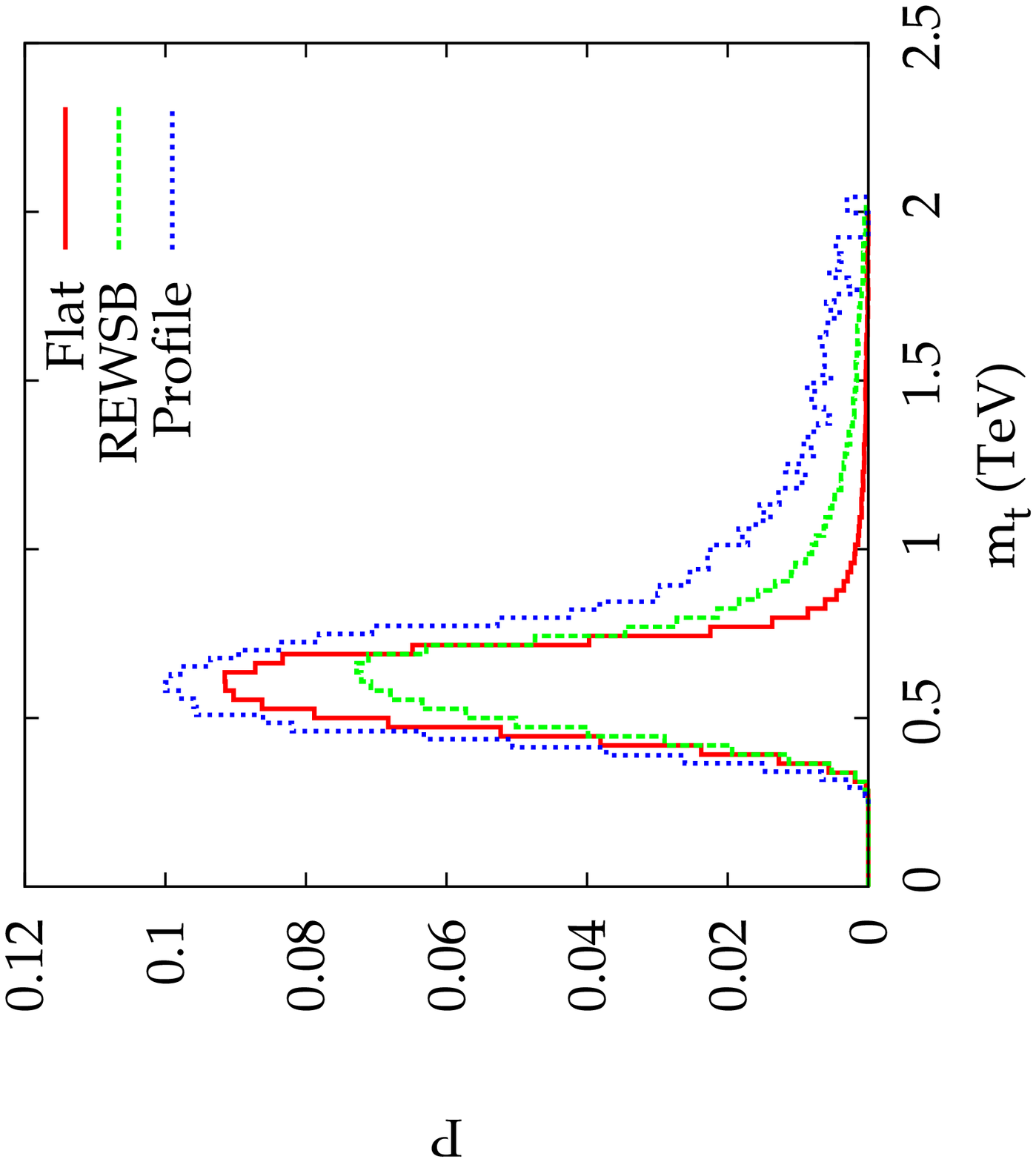}{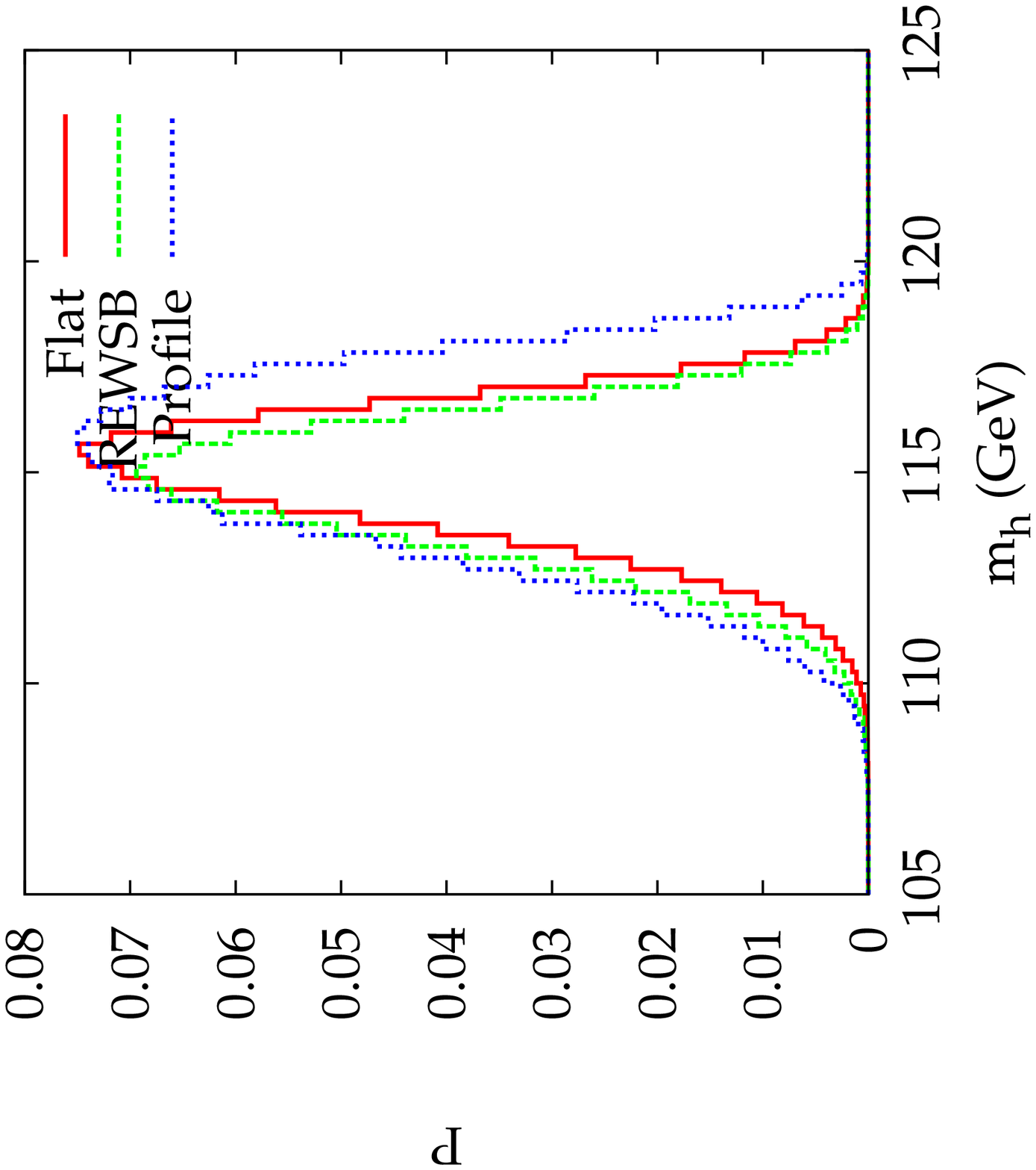}
\caption{1D Mass distributions for (a) the gluino $g$, (b) squark $\tilde{q}_{L}$, (c) selectron $\tilde{e}_R$ (d) the CP odd higgs $A^0$, (e) the lightest neutralino $\chi_1^0$, (f) the stop $\tilde{t}$ and (g) the lightest higgs $h$ for flat and REWSB priors with $\mu>0$, and with profile histograms. Profile likelihoods are discussed in Sect. \ref{sect:profile}, and have been rescaled to aid visual comparison.  \label{fig:masses}}}
\TABULAR{|c|c|c|}{\hline
Sparticle & Flat & REWSB  \\ \hline 
$g$ & 1216 & 1824 \\ 
$\tilde{q}_L$ & 1115 & 1655 \\
$\tilde{e}_R$ & 446 & 689 \\
$A^0$ & 846 & 1385  \\
$\chi_1^0$ & 426 & 669 \\
$\tilde{t}$ & 784 & 1189 \\ \hline }{95\% c.l. upper bounds for sparticle masses for $\mu>0$. All figures are in GeV. \label{tab:spartlim}}
Fig.~\ref{fig:masses} shows the likelihood distributions for the
masses of (a) the gluino, (b) the left-handed squark, (c) the
right-handed selectron, (d) the CP odd Higgs $A^0$, (e) the lightest
neutralino, (f) the stop and (g) the lightest CP even higgs $h$ for
$\mu>0$ with flat and REWSB priors, and for the profile likelihoods
discussed in Sect.~\ref{sect:profile}. Table \ref{tab:spartlim} shows
the 95\% confidence limit upper bounds on these masses. It is notable
that the 95\% upper bounds are almost all less than 1.5~TeV and
therefore lie well within the reach of the LHC within the next few
years. The higgs mass $m_h$ is constrained to lie below 120 GeV,
implying a late discovery of this particle at LHC. The priors have
only a small effect on the shape of the distribution, and our
predictions can therefore be considered quite robust.  

The main effect of the REWSB priors is to lengthen the tail out to
higher masses. This is because the REWSB prior favours lower
$\tan\beta$ which pushes the favoured region of parameter space up the
tail of the ``boomerang'' in Fig.~\ref{fig:m0tb}(b) to higher $m_0$
(and therefore higher $m_{1/2}$), thereby increasing the likelihood of
higher sparticle masses. Also of note is that the peaks of all three
distributions occur very close to one another. This independence of
the posterior pdfs from the priors and agreement with the profile
likelihood indicates that there are enough
observables constraining our model to overcome whatever prior beliefs
we might have. The predicted masses are therefore indicative of the
LVS and will not change with addition of more data. 

It is also interesting to quantify which observables are constraining the likelihood the most. An observable which is essentially constant over the parameter space will not constrain the posterior pdf very much, while one which exhibits significant variation over a range of a few standard deviations from the experimental central value would contribute a large $\chi^2$ to the likelihood in some regions but very little in others. To see how an individual observable is constraining the posterior pdf for the model variables $m_0$ and $\tan\beta$ we consider two cases: when it is used in the construction of the likelihood, and when it is omitted. To correctly take into account correlations between observables we consider the posterior pdf of $m_0$ jointly with $\tan\beta$, whose volume we normalise to one. We then calculate the integrated posterior difference of $\tan\beta$ jointly with $m_0$ which we call the ``moulding power'' 
\begin{equation}
M_P= 1/2 \int d\,\tan\beta d\, m_0 \left| p(\mbox{all data}| m_0 ,\tan\beta)- p(\mbox{all but one data}|m_0,\tan\beta)\right|.
\end{equation}
\TABULAR{|c|c|}{\hline
Observable & Constraint measure \\ \hline
$BR(B_s\to \mu\mu)$ & $0\pm 0$ \\
$\Delta M_{B_s}$ & $0 \pm 0$ \\
$\sin^2\theta_{\textup{eff}}$ & $0.007\pm0$\\
$BR(B\to \tau \nu)$ & $0.011 \pm 0$\\ 
$M_W$ & $0.051\pm0.001$ \\
$BR(b\to s \gamma)$ & $0.188\pm 0.003$ \\
$\Delta_{0-}$ & $0.208\pm0.006$ \\
$(g-2)_{\mu}$ & $0.390 \pm 0.012$ \\
$\Omega_{DM}h^2$ & $0.443\pm0.006$ \\
$m_h$ & $0.453 \pm 0.053$  \\
$\Omega_{DM}h^2$ (WMAP) & $0.799\pm0.005$  \\ \hline}{Moulding power of individual observables, as described in the text, for $\mu>0$. \label{tab:obsconstraints}} 
It is this quantity which we use as a measure of the effect an observable has on the likelihood, and the fact that each of the ten chains provides a statistically independent determination of $M_P$.  Table \ref{tab:obsconstraints} shows the estimates for $M_P$ obtained by this procedure. As expected $BR(B_s \to \mu\mu)$ and $\Delta M_{B_s}$ do not constrain the form of the likelihood. Similarly, $M_W$ and $\sin^2\theta_{\textup{eff}}$ are effectively constant on parameter space. The main constraining observables are the Higgs mass $m_h$ and the relic density $\Omega_{DM}h^2$, with the anomalous magnetic moment $(g-2)_{\mu}$ being the next most constraining.

We know that in mSUGRA that the unconstrained relic density can reach values as high as 100\cite{Allanach:2005kz}, and with the Gaussian
constraint we use, accurate sampling of the high relic density region is not possible for distributions with narrow allowed regions
and long tails. We found that the tail will was not accurately sampled, leading to bad statistics after reweighting resulting in a large value standard deviation of $M_P$. We therefore ran 10 chains of length 50000, omitting the relic density from the construction of the likelihood and used these chains to calculate $p(\mbox{all data but } \Omega_{DM}h^2|m_0,\tan\beta)$. In sharp contrast with mSUGRA we found no points where the relic density was higher than 1.5. For comparative purposes we have also calculated the constraint value for the relic density with the pure WMAP5 constraint in Table \ref{tab:obsconstraints}.

Finally, we  comment on the ``golden channel'' decay chain $\tilde{q}_L \to \chi^0_2 \to \tilde{l}_R \to \chi^0_1$. This important chain can give constraints on the mass spectrum \cite{spartmass} and even information about sparticle spins \cite{spartspin}. To calculate the probability of this chain existing in the LVS we find the fraction of points which have the mass ordering $m_{\tilde{q}_L} > m_{\chi^0_2} > m_{\tilde{l}_R} > m_{\chi^0_1}$. We find that this ordering occurs in all the points we generate, so that this chain should be useful for analysing LHC data.

\subsection{The Dark Side}
We now discuss the possibility that $\mu<0$. It is well known that there is a correlation between the sign of $\mu$ and the sign of the SUSY contribution to the anomalous magnetic moment of the muon in
constrained models like mSUGRA and the LVS (although not in the general MSSM). Hence while experiment currently favours $\mu>0$, and
it is true that a negative $\delta a_{\mu}$ leads to a large $\chi^2$ value, it could be possible to offset the bad fit to $\delta a_{\mu}$
with particularly good fits to other observables. Also of interest is the fact that it is possible to satisfy the $B$-term condition in
eq.~(\ref{eq:softterms}) for some limited regions of parameter space if $\mu<0$ and $m_s \sim 10^{14}\mbox{GeV}$. When this condition is
satisfied the LSP is the stau, which is ruled out by anomalous isotope abundance constraints. However it is conceivable that when the flux terms are computed and taken into account that this might change, and it might be possible to satisfy all the equations in (\ref{eq:softterms}) in a phenomenologically viable way, without introducing extra TeV scale matter as discussed above. We therefore press on in our exploration of the dark side of the Large Volume Scenario. 

\FIGURE{
\twographswcntrs{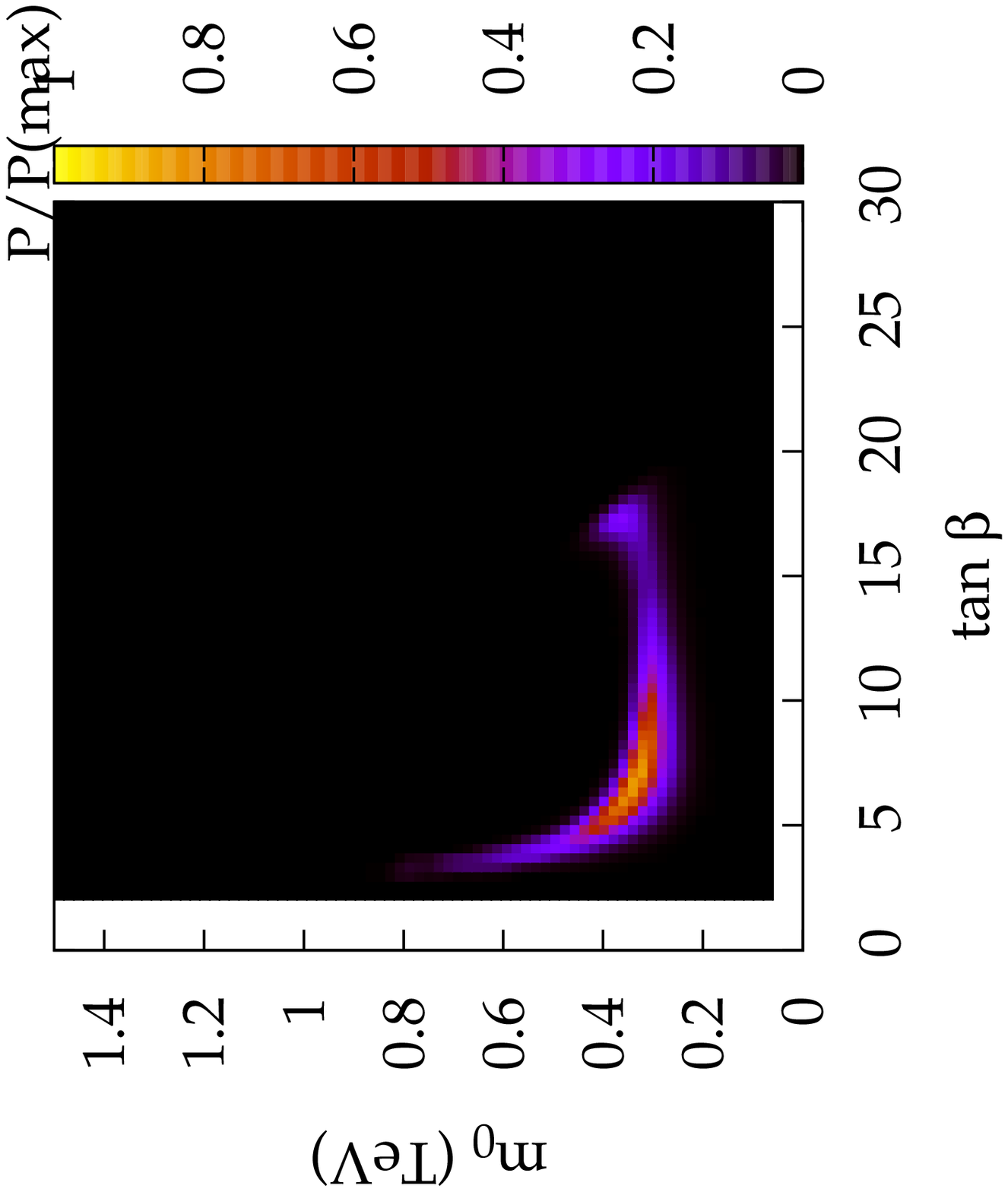}{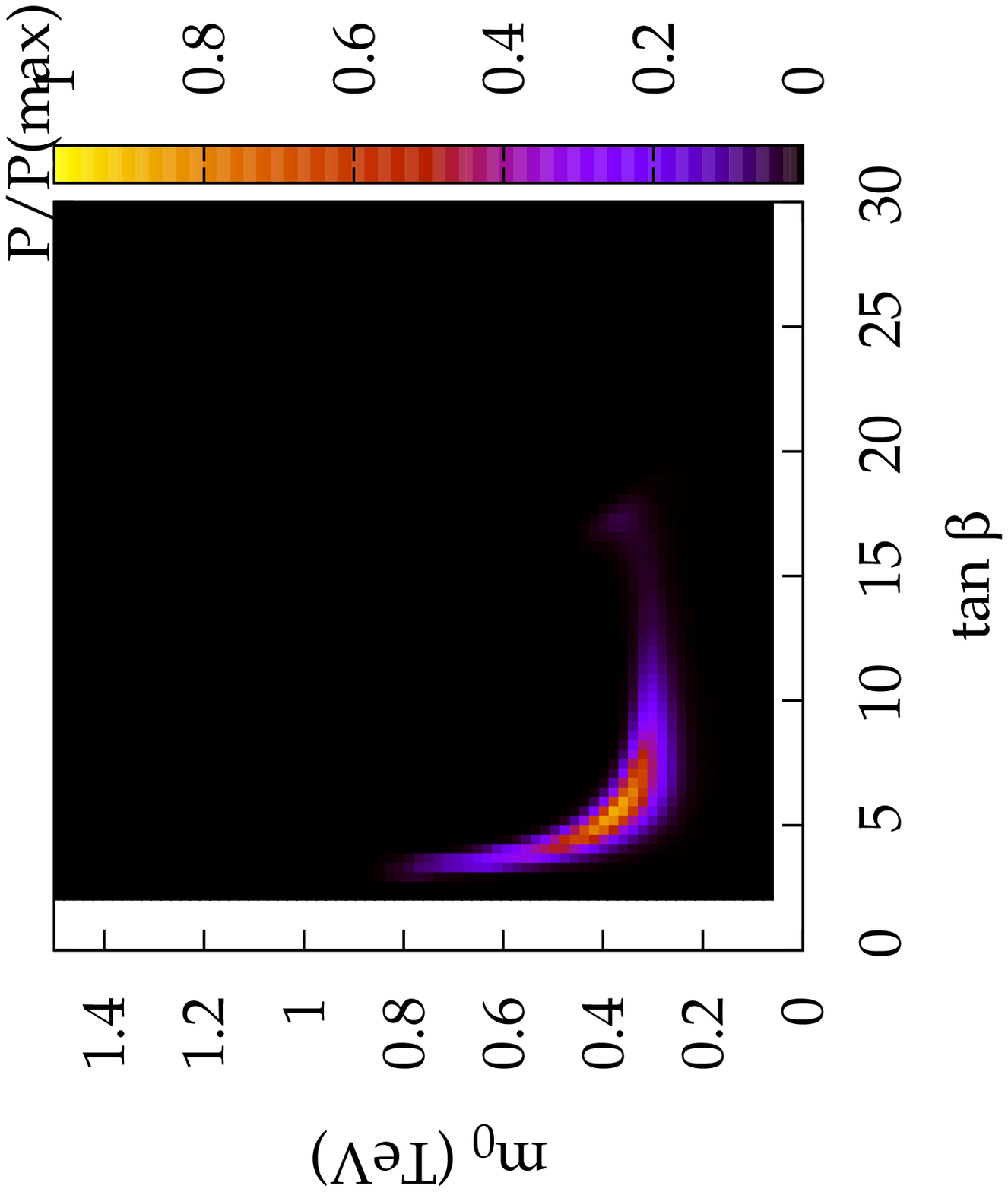}{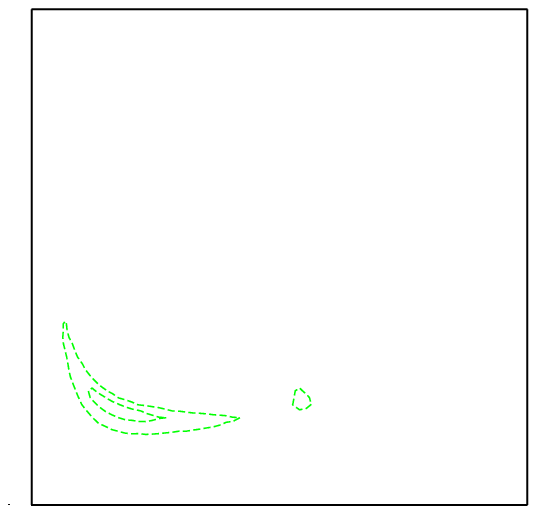}{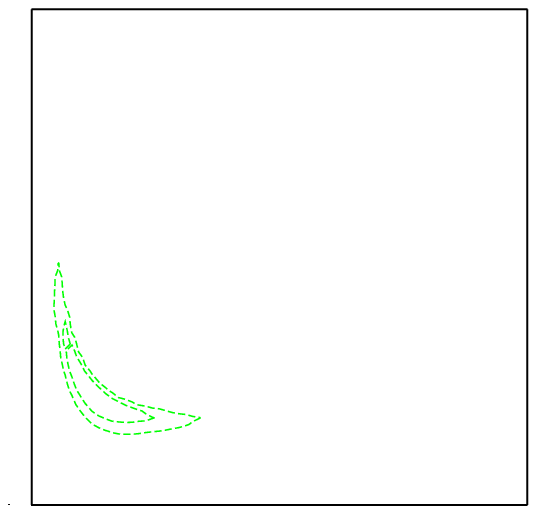}
\caption{Posterior pdfs for $\mu<0$ in the $m_0$-$\tan\beta$ plane for (a) flat priors and (b) REWSB priors with 68\% and 95\% c.l. contours shown. \label{fig:negmum0tb}}}

Fig.~\ref{fig:negmum0tb} shows the likelihood distributions marginalised to the $m_0$-$\tan\beta$ plane for both sets of priors with 68 and 95\% c.l. contours. The pdfs display a similar ``boomerang'' shape as Fig.~\ref{fig:m0tb} does in the  $\mu>0$ case, except the viable region of parameter space is considerably smaller. The posterior pdf for the flat priors exhibits some slight bimodality, which is eliminated by the REWSB priors' pull to lower $\tan\beta$. The favoured region in both cases is around the centre of the boomerang, with the REWSB priors increasing the amount of likelihood at higher $m_0$. Similar to the $\mu>0$ case, most of parameter space above and to the right of the favoured region is forbidden by having a stau LSP.

\TABULAR{|c|ccc|c|ccc|}{\hline
 & $\mu<0$  & $\mu>0$ &$\Delta \chi^2$ & & $\mu<0$ & $\mu>0$ & $\Delta
  \chi^2$ \\ \hline
$m_0$/GeV & 347.8  & 300.7&- &         $BR(b\to s\gamma)$ & 0.16 & 1.43& 0.05  \\
$\tan \beta$ &  7.1 & 14.7 &-     &    $BR(B_s\to\mu^+ \mu^-)$ & 0 & 0 & 0\\ 
$\Omega_{DM} h^2$ & 0.03 & 0&0.06 &    $\sin^2 \theta_{\textup{eff}}$ & 0.0 & 0.02 & 0.05  \\
$M_h$ & 1.17 & -0.48 & 0.25       &    $M_W$ & 0.77 & 0.94 & 0.08 \\
$(g-2)_{\mu}$ &  13.75 & 5.48 &0.31 &  $\Delta_{0-}$ & 1.26  & 3.55 & 0.09 \\ 
$\Delta M_{B_s}$ & 1.56 & 1.56 &0 &    $BR(B_u \to\tau\nu)$ &0.49  &0.57  &0.01 \\ \hline
&&&&                                   $\chi^2$ (total) & 20 & 13.6  & 1.41\\ \hline
}{Comparison of the best-fit points and statistical pulls for $\mu<0$ and $\mu>0$ with flat
  priors. $m_0$ and $\tan\beta$ are the values in parameter space, and
  all other values are $\chi^2$ values. The third column gives the
  error in our estimation of $\chi^2$, as described Section \ref{sec:fits}. \label{tab:bestfitnegmu}}
Table \ref{tab:bestfitnegmu} compares the best-fit points of both
signs of $\mu$, for flat priors. The WMAP upper bound is
saturated by the neutralino. The lightest Higgs mass falls just under
the $114.4\mbox{GeV}$ 95\% c.l. lower bound set by LEP2. $(g-2)_{\mu}$
is slightly negative, making a significant contribution to the overall
$\chi^2$. The $b$-observables $BR(b\to s\gamma)$ and $\Delta_{0-}$ are
fitted slightly better by $\mu<0$, while for $\Delta M_{B_s}$,
$\sin\theta_{\textup{eff}}$ and $M_W$ there is essentially no
difference. The branching ratio $BR(B\to\mu^+\mu^-)$ comfortably
evades experimental bounds in both cases. As in Section \ref{sec:fits} we present the error in our estimation of the $\chi^2$s of the observables, and the total error in $\chi^2$.

The overall $\chi^2$ of the fit is 20 and 20.29 for the flat and REWSB
priors respectively. In both cases nearly 65\% of the $\chi^2$ comes
from $(g-2)_{\mu}$. If we omit this observable, then the $\chi^2$
is 6.25 (7.69), which is roughly the same as for $\mu>0$
without $(g-2)_{\mu}$, which has $\chi^2=$ 8.10 (5.77). Therefore, if were to omit the anomalous magnetic moment from the likelihood both signs of $\mu$ would be consistent with the data. Calculating the $p$-values associated with the total $\chi^2$s (including $(g-2)_{\mu}$) we obtain $P_{\textup{flat}}(\mu<0)=0.010$ and $P_{\textup{REWSB}}(\mu<0)=0.009$. This is statistically significant down to the 2\% level, and allows us to reject the possibility that $\mu<0$.

\section{Profile Likelihoods}
\label{sect:profile}
\FIGURE{ \onegraphwcntrs{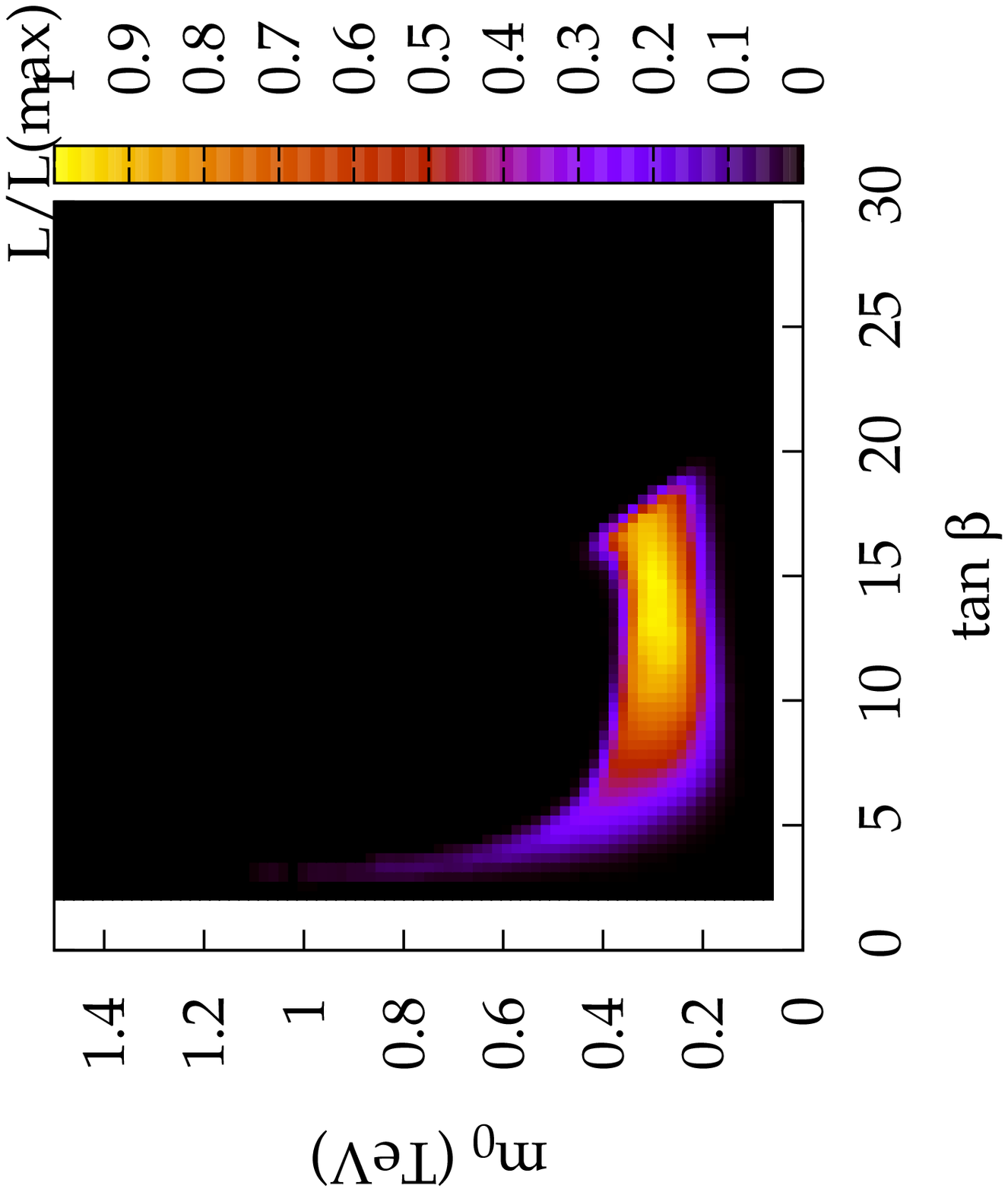}{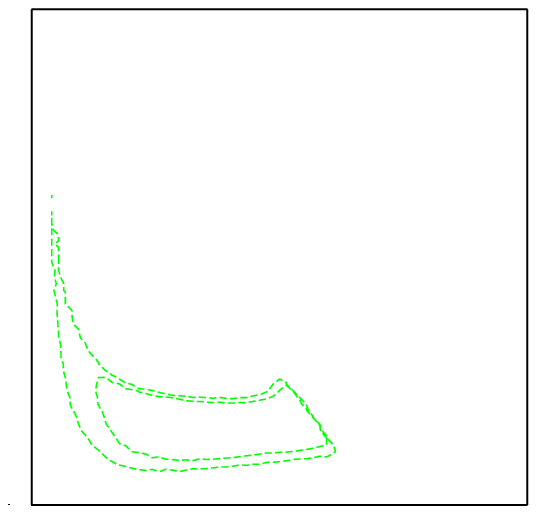} \caption{2D profile likelihoods in the $m_0$-$\tan\beta$ plane for $\mu>0$, including 70\% and 95\% confidence limit contours. \label{fig:profilem0tb}}}
A regular objection to the Bayesian analysis presented above comes in
the form of criticism about the the subjectivity of the
priors. While  we have shown that in the LVS the posterior pdfs are
essentially independent of the priors, we believe it fair to present a
frequentist analysis of the model.  Frequentists prefer to get rid of
nuisance parameters by maximising them. This is known as concentration
of parameters and the likelihood function of the reduced parameter set
is called the profile likelihood. However, it is important to state
that profile likelihood plots are not pdfs, as the profile likelihood
is not derived from a probability distribution. We can easily derive
the profile likelihood from the Markov chains we have generated as
follows \cite{naturalness}: we bin the chains in the usual way, and then find the maximum
likelihood in each bin and plot that. The 95\% and 70\%  c.l. regions are then defined by $2\Delta \ln L = 5.99$ (2.41) respectively where $\Delta\ln L = \ln L_{max} -\ln L$ \cite{minuit}.

In Fig.~\ref{fig:profilem0tb} we plot the profile likelihood with 70\% and 95\% c.l. contours in the $m_0$-$\tan\beta$ plane. The plot is reassuringly similar to those in Fig.~\ref{fig:m0tb}. This illustrates an important point: that given enough data we expect the profile likelihood and the Bayesian likelihoods to look the same. Fig.~\ref{fig:masses} shows profile histograms for some relevant sparticles. The profile histograms have been rescaled to ease comparison with the Bayesian posteriors on the same plots. The profile results are in good agreement with the Bayesian histograms, further illustrating the point made above regarding convergence of profile and Bayesian likelihoods in the presence of ample data. The tails of the histograms are slightly noisier than those of the Bayesian posteriors, but this could be eliminated by running the MCMCs for longer.

Using profile likelihoods we can establish the nature of the spike feature discussed in Sect.~\ref{sec:fits}. We show in Fig.~\ref{fig:profiledm}
the profile likelihoods for $\Omega_{DM}h^2$ and for the mass
splitting $m_{\tilde{\tau}}- m_{\chi^0_1}$, which have been multiplied
by constants
for comparison with the  Bayesian posterior distribution with flat
priors. Fig.~\ref{fig:profiledm}(a) shows that the region of very low
relic density does not fit the data any better than other regions of
parameter space. This establishes that the spike in the relic density
is a volume effect, as promised earlier. In
Fig.~\ref{fig:profiledm}(b) we see that the peak in the
stau-neutralino mass splitting is also a volume effect, which was
shown in Fig.~\ref{fig:chistau} to be associated with the spike. 
\FIGURE{\twographs{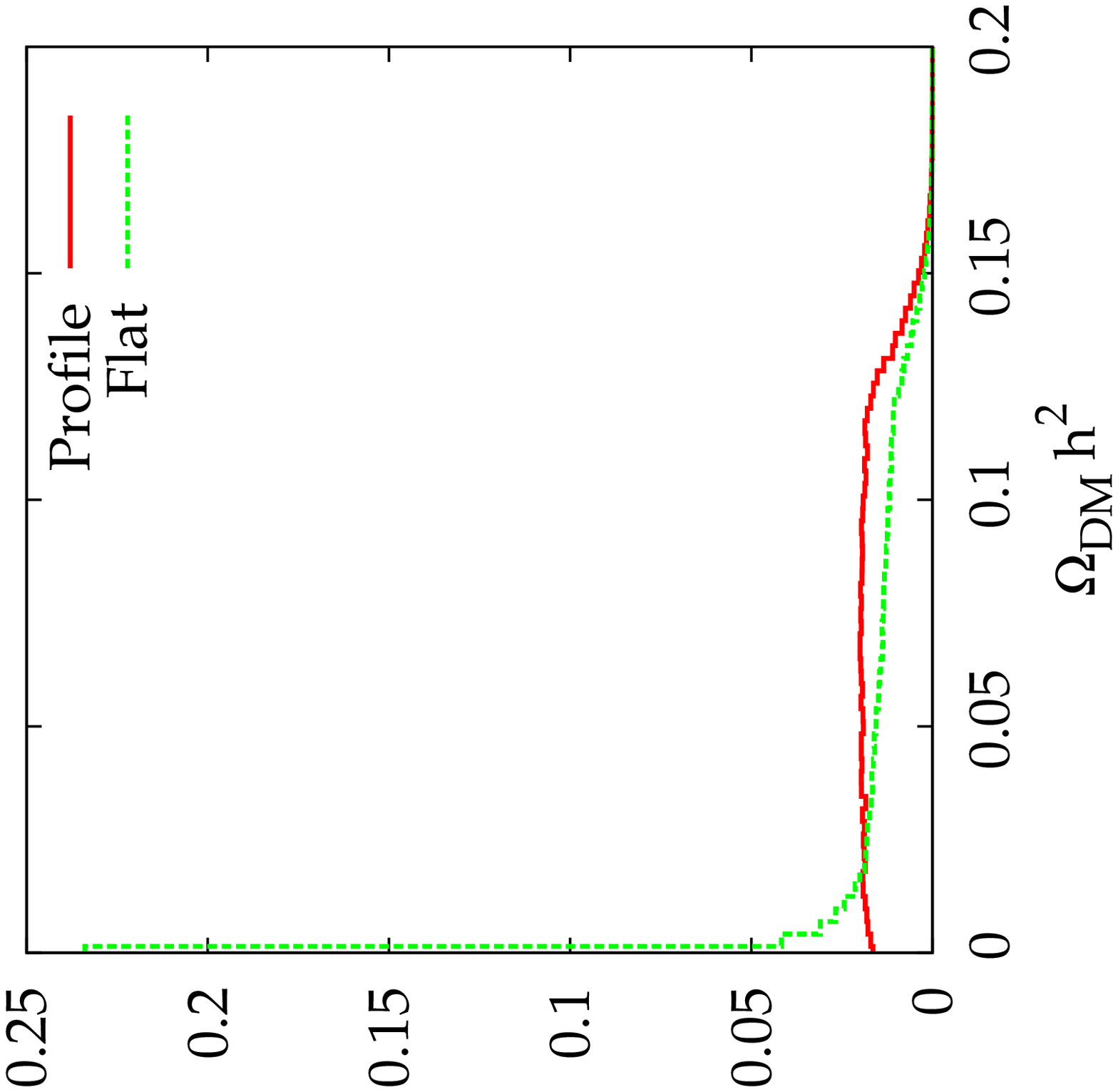}{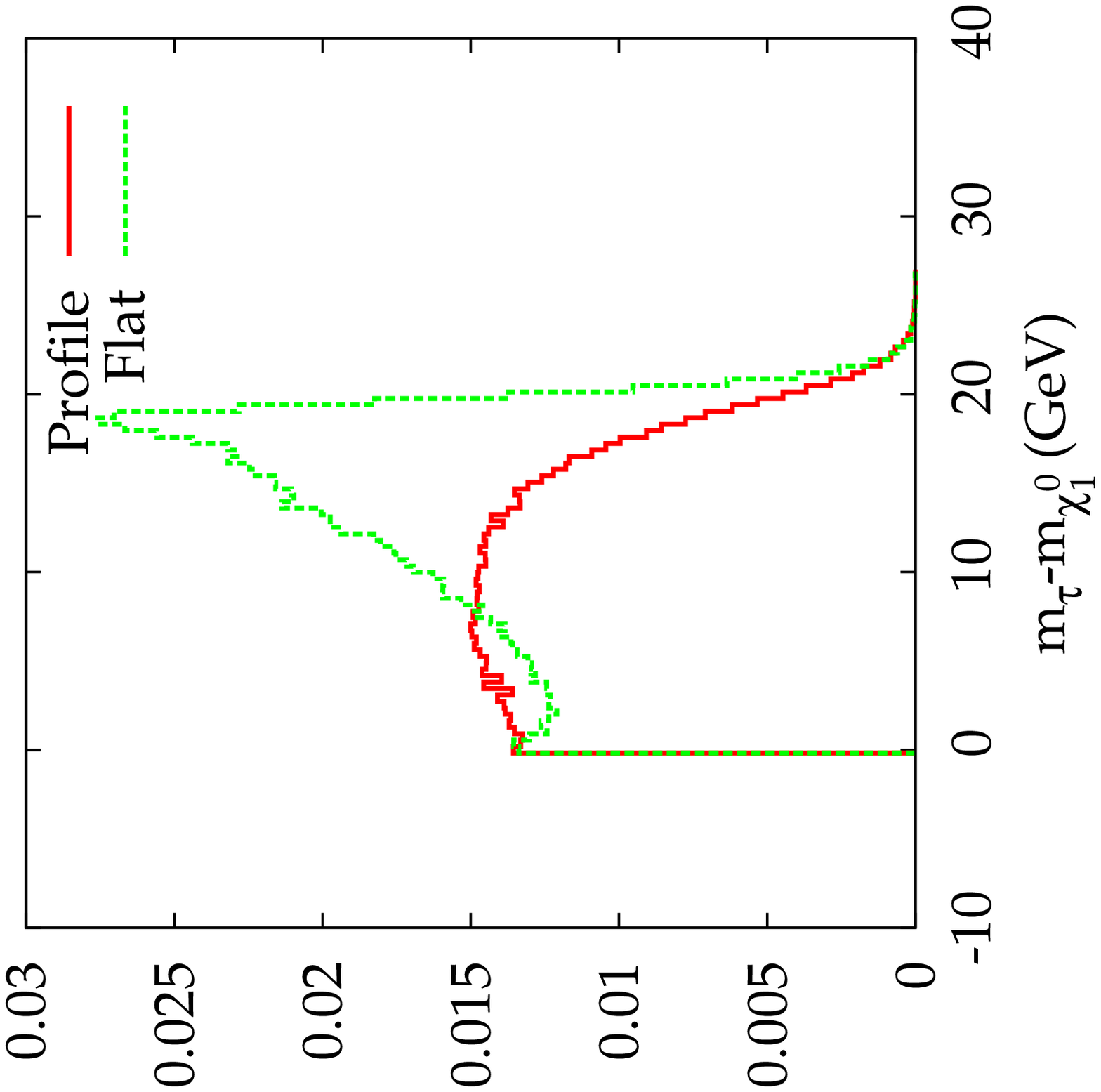}
  \caption{Histograms for (a) dark matter relic density and (b)
    $m_{\tilde{\tau}}-m_{\chi^0_1}$ for $\mu>0$. For the  ``flat''
    plot we show the posterior pdf per bin. For the ``profile'' plot
    we show the likelihood profile discussed in the text. To aid
    visual comparison the profile plots have been
    multiplied by constants. \label{fig:profiledm}}}

\section{Conclusion}
We have used  Monte Carlo Markov chain methods to make global fits to the Large Volume Scenario in the minimal case with modular parameter $\lambda=1/3$, the first time this method has been applied to a model derived directly from string theory. As indirect constraints on the model we have used the WMAP 5-year dataset, the most recent measurement of $m_t$ from the Tevatron and a suite of other electroweak and B-observables sensitive to flavour changing neutral currents, including for the first time in MCMC fits $BR(B\to\tau\nu)$, $\Delta M_{B_s}$ and the isospin asymmetry $\Delta_{0-}$. We have shown that the model is constrained enough that the choice of flat or REWSB priors does not make a radical difference to the posterior pdf, unlike the CMSSM case. This illustrates the point that although the priors encode our uncertainty or \textit{a priori} beliefs about a quantity, given enough data nature will speak for herself and the posterior pdfs will become essentially independent of the priors. Furthermore, the frequentist profile likelihoods we have presented have identified the same region of parameter space as the Bayesian likelihood, indicating the robustness of our fits which again contrasts with the CMSSM case. We have constructed a new quantitative measure to determine which observables are constraining the form of the likelihood the most, based on examining the difference between posterior pdfs when we include and omit an observable from the construction of the likelihood. We find that $(g-2)_{\mu}$, the Higgs mass and the dark matter relic density have the greatest effect on the fits.

We have also investigated both signs of $\mu$, and by calculating P-values from the $\chi^2$ of the best-fit points in both cases we have rejected the possibility that $\mu<0$, while $\mu>0$ is consistent with our constraints.

One of the main contributions of this paper is in forecasting what might be seen at the LHC. We find that \textit{all} the points we generate allow the ``golden channel'' cascade $\tilde{q}_L \to \chi^0_2 \to \tilde{l}_R \to \chi^0_1 $. Use of this cascade should allow us to extract much information about masses and even spins of observed sparticles. Also, the 95\% upper bounds on several important sparticle masses all fall below 1.2 TeV within reach of the LHC, implying early SUSY discovery at LHC. However the probability that $m_{\tilde{\tau}}-m_{\chi^0_1} <10$GeV is 38\% so that reconstructing the stau may prove difficult, and the mass of the lightest Higgs $m_h$ is constrained to be less than $120\mbox{GeV}$ indicating that the Higgs may not be found at LHC for a number of years. A ``smoking gun'' signature for discriminating the model against the CMSSM could be the ratio of gaugino masses $1.5-2:2:6$ for the LVS, compared to $1:2:6$ in the CMSSM, as first noticed in ref.\cite{LVSLHC}.  In our model $\tan\beta$ is bounded from above at around $\tan\beta= 20$, in contrast with the CMSSM where the region with the highest likelihood occurs where $\tan\beta>50$. While we must take into account the inevitable fuzziness of our bounds due the fluxes, a precise measurement of $\tan\beta$, although difficult, could therefore prove sufficient to reject the minimal LVS.

Further work in this vein could include taking into account the modular parameter $\lambda$ as an extra variable in parameter space, thereby indirectly probing the geometry of the Calabi-Yau on which the model is compactified. In terms of connecting with the LHC, now that we have successfully sampled the likelihood distribution it is possible to construct correctly weighted samples of collider observables, as done using randomly generated points in \cite{LVSLHC}. To escape from the dilute flux approximation it is ultimately desirable to take into account analytically the fluxes on the Calabi-Yau, thereby allowing fully realistic behaviour at the string scale. This is unfortunately easier said than done, but the effects this would have on string phenomenology should not be underestimated.

\section*{Acknowledgements}  We would like to thank J.P. Conlon and  F. Quevedo for useful discussions regarding the LVS, and S. AbdusSalam and S. Meinel for help with B observables. MJD is supported by EPSRC, the Cambridge European Trusts and St. John's College, Cambridge. These simulations were run on the Camgrid cluster with the help of Mark Calleja. This work is partially supported by STFC.


\begin{thebibliography}{99}
\bibitem{unistring}
J.~P.~Conlon and F.~Quevedo, 
{\em Gaugino and scalar masses in the landscape},
JHEP {\bf 0606} (2006) 029,
[arXiv:hep-th/0605141];
L.~E.~Ibanez, 
{\em The fluxed MSSM},
Phys.~Rev.~D {\bf 71} (2005) 055005,
[arXiv:hep-ph0408064];
A.~Brignole, L.~E.~Iba\~nez and C.~Mu\~noz,
{\em Towards a theory of soft-terms for the supersymmetric Standard Model}
Nucl.~Phys.~B {\bf 422} (1994) 125,
Erratum-idis B {\bf436} (1995) 747,
[arXiv:hep-ph9308271]
\bibitem{LVS}
V. Balasubramanian, P.~Berglund, J.~P.~Conlon and F.~Quevedo,
{\em Systematics of Moduli Stabilisation in Calabi-Yau Flux Compactifications},
JHEP {\bf 0503} (2005) 007,
[arXiv:hep-th/0502058]
\bibitem{intermediate}
C.~P.~Burgess, L.~E.~Iba\~nez and F.~Quevedo,
{\em Strings at the Intermediate Scale, or is the Fermi Scale Dual to the Planck Scale?},
Phys.~Lett.~ {\bf B447} (1999) 257,
[arXiv:hep-ph/9810535]
\bibitem{KKLT}
S.~Kachru, R.~Kallosh, A.~Linde and S.~Trivedi,
{\em de Sitter Vacua in String Theory},
Phys.Rev. {\bf D68} (2003) 046005,
[arXiv:hep-th/0301240]
\bibitem{LVSsoftterms}
J.~P.~Conlon, S.~S.~Abdussalam, F.~Quevedo and K.~Suruliz,
{\em Soft SUSY Breaking Terms for Chiral Matter in IIB String Compactifications},
JHEP {\bf 0701} (2007) 022,
[arXiv:hep-th/0610129]
\bibitem{LVSLHC}
J.P.~Conlon {\em et al.},
{\em Sparticle Spectra and LHC signatures for Large Volume String Compactifications},
JHEP {\bf 08} (2007) 061,
[arXiv:0704.3403]
\bibitem{Allanach:2005kz}
  B.~C.~Allanach and C.~G.~Lester,
  {\em Multi-dimensional mSUGRA likelihood maps},
  Phys.\ Rev.\ D {\bf 73} (2006) 015013
  [arXiv:hep-ph/0507283].
\bibitem{naturalness}
B. C. Allanach, K. Cranmer, C. G. Lester and A. M. Weber,
{\em Natural Priors CMSSM Fits and LHC Weather Forecasts},
JHEP {\bf 08} (2007) 023,
[arXiv: 0704.0487]
\bibitem{deAustri:2006pe}
  R.~R.~de Austri, R.~Trotta and L.~Roszkowski,
  {\em A Markov chain Monte Carlo analysis of the CMSSM},
  JHEP {\bf 0605} (2006) 002
  [arXiv:hep-ph/0602028]
\bibitem{leszek}
 R.~R.~de Austri, R.~Trotta and L.~Roszkowski,
{\em On the detectability of the CMSSM light Higgs boson at the Tevatron},
JHEP {\bf 0704} (2007) 084,
[arXiv:hep-ph/0611173];
{\em Implications for the Constrained MSSM from a new prediction for b to s gamma}
JHEP {\bf0707} (2007) 075, 
[arXiv:0705.2012];
{\em On prospects for dark matter indirect detection in the Constrained MSSM}
[0707.0622]
\bibitem{bayes} T.~Bayes,
{\em A letter to John Canton},
Phil.~Trans.~Royal Society London {\bf 53}: 269-71 (1763).
\bibitem{weiglein}
S.~Heinemeyer, X.~Miao, S.~Su and G.~Weiglein,
{\em B-Physics Observables and Electroweak Precision Data in the CMSSM, mGMSB and mAMSB},
[arXiv:0805.2359]


\bibitem{inflation}
J.~P.~Conlon and F.~Quevedo,
{\em K\"ahler Moduli Inflation},
JHEP {\bf 0601} (2006) 146,
[arXiv:hep-th/0509012]
\bibitem{axion}
J.~P.~Conlon,
{\em The QCD Axion and Moduli Stabilisation}
JHEP {\bf 0605} (2006) 078,
[arXiv:hep-th/0602233]
\bibitem{mneut}
J.~P.~Conlon and D.~Cremades,
{\em The Neutrino Suppression Scale from Large Volumes}
P.R.L. {\bf 99} 041803 (2007)
[arXiv:hep-ph/0611144]
\bibitem{joethesis}
J.~P.~Conlon,
{\em Moduli Stabilisation and Applications in IIB String Theory},
[arXiv:hep-th/0611039]
\bibitem{dineseiberg}
M.~Dine and N.~Seiberg,
{\em Is the Superstring Weakly Coupled?},
Phys.~Lett. {\bf B162} (1985) 299
\bibitem{loopjump}
M.~Berg, M.~Haack and E.~Pajer,
{\em Jumping Through Loops: On Soft Terms from Large Volume Compactifications},
JHEP {\bf 0709} (2007) 031,
[arXiv:0704.0737]
\bibitem{michele}
M.~Cicoli, J.~P.~Conlon and F.~Quevedo,
{\em Systematics of String Loop Corrections in Type IIB Calabi-Yau Flux Compactifications},
JHEP {\bf 0801} (2008) 052,
[arXiv:0708.1873]
\bibitem{torus}
D.~L\"ust, P.~Mayr, R.~Richter and S.~Stieberger,
{\em Scattering of Gauge, Matter and Moduli Fields from Intersecting Branes},
Nucl.~Phys. {\bf B696} (2004) 205,
[arXiv:hep-th/0404134]
\bibitem{brignole}
A.~Brignole, L.~E.~Iba\~nez and C.~Mu\~noz,
{\em Soft supersymmetry breaking terms from supergravity and superstring models},
[arXiv:hep-ph/9707209]
\bibitem{kahler}
J.~P.~Conlon, D.~Cremades and F.~Quevedo,
{\em K\"ahler Potentials of Chiral Matter Fields for Calabi-Tai String Compactifications},
JHEP {\bf 0701} (2007) 022,
[arXiv:hep-th/0609180]


\bibitem{Allanach:2006jc}
  B.~C.~Allanach,
   {\em Naturalness priors and fits to the constrained minimal supersymmetric
  standard model},
  Phys.\ Lett.\ B {\bf 635} (2006) 123
  [arXiv:hep-ph/0601089]
\bibitem{pdg}
W.~M.~Yao \textit{et al.} (Particle Data Group),
J.~Phys. {\bf G33} (2006) 1 and 2007 partial update for 2008 edition.
\bibitem{alphas}
T.~Becher and M.~D.~Schwarz,
{\em A precise determination of $alpha_s$ from LEP thrust data using effective field theory},
arXiv:0803.0342
\bibitem{mt}
Tevatron Electroweak Working Group,
{\em A Combination of CDF and D0 Results on the Mass of the Top Quark},
arXiv:0803.1683
\bibitem{softsusy}
B.C. Allanach, {\em SOFTSUSY: A program for calculating supersymmetric
  spectra}, Comput. Phys. Commun. {\bf 143} (2002) 305, [arXiv:hep-ph/0104145].
\bibitem{slha}
P. Skands {\em et al}, {\em SUSY Les Houches accord: Interfacing SUSY spectrum
  calculators, decay packages, and event generators}, JHEP {\bf 0407} (2004)
036, [arXiv:hep-ph/0311123].  
\bibitem{micromegas}
 G. B\'{e}langer, F. Boudjema, A. Pukhov and A. Semenov,
{\em micrOMEGAs: Version 1.3},
 Comput. Phys. Commun. {\bf 174} (2006) 577 [arXiv:hep-ph/0405253];
G. B\'{e}langer, F. Boudjema, A. Pukhov and A. Semenov,
{\em micrOMEGAs: A program for calculating the relic density in the MSSM},
 Comput. Phys. Commun. {\bf 149} (2002) 103 [arXiv:hep-ph/0112278].  
 \bibitem{superiso}
 F.~Mahmoudi,
 {\em SuperIso: A program for calculating the isospin asymmetry of $B\rightarrow K^* \gamma$ in the MSSM},
Comput.~Phys.~Commun. {\bf 178} (2008) 745,
 [arXiv:0710.2067]
\bibitem{WMAP}
E.~Komatsu \textit{et al.},
{\em Five-Year Wilkinson Microwave Anisotropy Probe (WMAP) Observations: Cosmological Interpretation},
arXiv:0803.0547
\bibitem{omegaconstraint}
A.~Arbey and F.~Mahmoudi,
{\em SUSY constraints from relic density: high sensitivity to pre-BBN expansion rate},
[arXiv: 0803.0741]
\bibitem{RHNDM}
V.~Barger, D.~Marfatia and A.~Mustafayev,
{\em Neutrino Sector impacts SUSY dark matter},
[arXiv:0804.3601]
\bibitem{SMmuon}
J.~P.~Miller, E.~de Rafael and B.~L.~Roberts,
{\em Muon g-2: Review of Theory and Experiment},
Rept.\ Prog.\ Phys.\ {\bf 70} (2007) 795-881,
 [arXiv:hep-ph/0703049]
\bibitem{susycont}
 U.~Chattopadhyay and P.~Nath,
  {\em Probing supergravity grand unification in the Brookhaven g-2 experiment},
  Phys.\ Rev.\ D {\bf 53} (1996) 1648,
  [arXiv:hep-ph/9507386].
\bibitem{2loopg-2}
S.~Heinemeyer, D.~St\"ockinger, G.~Weiglein,
{\em Electroweak and supersymmetric two-loop corrections to $(g-2)_{\mu}$},
Nucl.\ Phys.\ B{\bf 690} (2004) 103,
[arXiv:hep-ph 0405255];
S.~Heinemeyer, D.~St\"ockinger, G.~Weiglein,
{\em Two loop SUSY corrections to the anomalous magnetic moment of the muon},
Nucl.\ Phys.\ B{\bf 690} (2004) 62,
[arXiv: hep-ph/0312264]
\bibitem{susymuon}
D.~St\"ockinger,
{\em The muon magnetic moment and supersymmetry},
J.\ Phys.\ G{\bf34} (2007) R45-R92,
[arXiv:hep-ph/0609168]  
\bibitem{wmass}
The LEP Collaborations,
{\em Precision Electroweak Measurements and Constraints on the Standard Model},
arXiv:0712.0929.
\bibitem{Heinemeyer:2006px}
  The code is forthcoming in a publication by A.~M.~Weber et al.;
  S.~Heinemeyer, W.~Hollik, D.~St\"ockinger, A.~M.~Weber and G.~Weiglein,
  {\em Precise prediction for M(W) in the MSSM},
JHEP {\bf 08} (2006) 052,
  [arXiv:hep-ph/0604147].
\bibitem{Mwerror}
  S.~Heinemeyer, W.~Hollik, A.M.~Weber, G.~Weiglein,
  {\em Z Pole Observables in the MSSM},
  arXiv:0710.2972.
\bibitem{bsgamma}
Heavy Flavour Averaging Group,
{\em Averages of b-hadron properties at the end of 2006},
[arXiv:0704.3575]  
\bibitem{nazila0710}
F.~Mahmoudi,
{\em New constraints on supersymmetric models from $b \to s \gamma$},
JHEP {\bf 12} (2007) 026,
[arXiv: 0710.3791]
 \bibitem{bsmumu}
  CDF Collaboration,
  {\em Search for $B_s \to \mu^+ \mu^-$ and $B_d \to \mu^+\mu^-$ Decays with 2$\mbox{fb}^{-1}$ of ppbar Collisions},
  arXiv: 0712.1708.  
\bibitem{hfagbtnu}
Heavy Flavour Averaging Group,
available at {http://www.slac.stanford.edu/xorg/hfag/rare/leppho07/radll/index.html}  
\bibitem{btnu1}
UTFit Collaboration,
available at {http://utfit.roma1.infn.it/summer2006/ckm-results/ckm-results.html}
\bibitem{isidori2006}
 G.~Isidori \& P.~Paradisi,
{\em Hints of large $\tan\beta$ in flavour physics},
Phys.~Lett.~ {\bf B639} (2006) 499,
[arXiv:hep-ph/0605012]
\bibitem{buras}
A.~J.~Buras, P.~H.~Chankowski, J.~Rosiek and \L.~S\l awianowska,
{\em $\Delta M_{d,s}$, $B^0_{d,s}\to\mu^+\mu^-$ and $B\to X_s\gamma$ in Supersymmetry at Large $\tan\beta$},
Nucl.~Phys.~ {\bf B659} (2003) 3,
[arXiv:hep-ph/0210145]
\bibitem{deltamsexp}
CDF Collaboration,
  {\em Observation of Bs-Bsbar Oscillations},
  Phys.\ Rev.\ Lett.\ {\bf 97} (2006) 242003,
  [arXiv:hep-ex/0609040].
\bibitem{UTFit}
M. Bona { \em et al.} (UTFit Collaboration),
{\em The Unitarity Triangle Fit in the Standard Model and Hadronic Parameters from Lattice QCD: A Reappraisal after the Measurements of $\Delta m_s$    and BR($B \rightarrow \tau \nu$)},
JHEP {\bf 0610} (2006) 081,
[arXiv: hep-ph/0606167]
\bibitem{S0}
G.~Buchalla, A.~J.~Buras and M.~E.~Lautenbacher,
{\em Weak Decays Beyond Leading Logarithms}
Rev.~Mod.~Phys.~ {\bf68} (1996) 1125,
[arXiv:hep-ph/9512380]
\bibitem{wingate}
A.~Gray et al.,
{\emph The B Meson Decay Constant from Unquenched Lattice QCD},
Phys.~Rev.~Lett.~{\bf 95} (2005) 212001, 
[arXiv:hep-lat/0507015];
A.~Gray et al.,
{\emph The Upsilon spectrum and $m_b$ from full lattice QCD},
Phys.~Rev.~ {\bf D72} (2005) 094507,
[arXiv:hep-lat/0507013]
\bibitem{isosymbabar} 
 B. Aubert {\em et al.} [BABAR Collaboration],
 {\em Measurement of Branching Fractions, and CP and Isospin Asymmetries, for $B \to K^* \gamma$},
 Phys.\ Rev.\  D{\bf 70} (2004) 112006,
 [arXiv: hep-ex/0407003].
\bibitem{isosymbelle}
M. Nakao {\em et al.} [Belle Collaboration],
{\em Measurement of the $B \to K^* \gamma$ Branching Fractions and Asymmetries}, 
	Phys.\ Rev.\ D{\bf 69} (2004) 112001,
	[arXiv: hep-ex/0402042].
\bibitem{cslin}
Thanks to C.S. Lin for the likelihood.
\bibitem{higgs}
B.C. Allanach {\em et al.},
{\em Precise determination of the neutral Higgs boson masses in the MSSM},
JHEP {\bf 0409} (2004) 044,
[arXiv:hep-ph/0406166]
\bibitem{ewsb}
D. M. Pierce, J. A. Bagger, K. T. Matchev and R.-J Zhang,
{\em Precision corrections in the minimal supersymmetric standard model},
Nucl.\ Phys.\ {\bf B491} (1997) 3-67,
[arXiv: hep-ph/9606211]
\bibitem{gelmanrubin}
A. Gelman and D. Rubin,
{\em Inference from Iterative Simulation Using Multiple Sequences},
Stat.\ Sci.\ {\bf 7} (1992) 457.
\bibitem{susyspace}
J.~Ellis, S.~Heinemeyer, K.A.~Olive, A.M.~Weber \& G.~Weiglein
{\emph The Supersymmetric Parameter Space in Light of B-physics Observables and Electroweak Precision Data},
[arXiv:0706.0652].
\bibitem{spartmass}
B.~C.~Allanach, C.~G.~Lester, M.~A.~Parker and B.~R.~Webber,
{\em Measuring sparticle masses in non-universal string inspired models at the LHC},
JHEP {\bf 0009} (2000) 004,
[arXiv:hep-ph/0007009]
\bibitem{spartspin}
A.~B.~Barr,
{\em Using lepton charge asymmetry to investigate the spin of supersymmetric particles at the LHC},
Phys.~Lett.~B {\bf 596} (2004) 205,
[arXiv:hep-ph/0405052]
\bibitem{minuit}
F.~James,
{\em Minuit, function minimization and error analysis},
CERN Program Library Long Writeup D506, Section 7.3.
\end{thebibliography}
\end{document}